\documentclass[twocolumn,aps,prl,showpacs,tightenlines]{revtex4-1}
\pdfoutput=1
\usepackage{amsmath}
\usepackage{amsfonts}
\usepackage{graphicx}
\usepackage{epsfig}
\usepackage{color}
\usepackage[colorlinks,citecolor=blue]{hyperref}

\begin{document}

\title{Macroscopic Quantum Superposition in Cavity Optomechanics}
\author{Jie-Qiao Liao}
\email{jieqiaoliao@gmail.com}
\affiliation{School of Natural Sciences, University of California, Merced, California 95343, USA}
\author{Lin Tian}
\email{ltian@ucmerced.edu}
\affiliation{School of Natural Sciences, University of California, Merced, California 95343, USA}
\date{\today}

\begin{abstract}
Quantum superposition in mechanical systems is not only key evidence for macroscopic quantum coherence, but can also be utilized in modern quantum technology. Here we propose an efficient approach for creating macroscopically distinct mechanical superposition states in a two-mode optomechanical system. Photon hopping between the two cavity modes is modulated sinusoidally. The modulated photon tunneling enables an ultrastrong radiation-pressure force acting on the mechanical resonator, and hence significantly increases the mechanical displacement induced by a single photon. We study systematically the generation of the Yurke-Stoler-like states in the presence of system dissipations. We also discuss the experimental implementation of this scheme.
\end{abstract}

\maketitle

\emph{Introduction.}---Quantum superposition~\cite{Dirac1958} is at the heart of quantum theory and is often considered a signature to distinguish the quantum from the classical world. To date quantum superposition has been observed in various physical systems~\cite{Arndt2014}, such as electronic~\cite{Clarke1988,Friedman2000,Mooij2000}, photonic~\cite{Brune1996,Polzik2006,Ourjoumtsev2006b,Schoelkopf2013}, and atomic or molecular systems~\cite{Monroe1996,Zeilinger1999}, ranging from microscopic systems to mesoscopic devices.  Nevertheless, it would be desirable to observe quantum superposition in macroscopic mechanical systems with up to $10^{10}$ atoms~\cite{Leggett2002}. It can help us understand the fundamentals of quantum theory~\cite{Zurek1991}, such as quantum decoherence and quantum-classical boundary in the presence of gravity~\cite{Blencowe2013}, and has wide applications in quantum information processing with continuous variables~\cite{Schoelkopf2013}.

Recent advances in microfabrication provide the possibility of producing high-$Q$ mechanical resonators~\cite{Schwab2005}. This progress paves the way for observing and utilizing quantum effects in macrosized mechanical systems~\cite{Rugar2004,Naik2006,Regal2008,Cleland2010,Teufel2011,Chan2011,Kippenberg2012,Kolkowitz2012}. Great efforts have been devoted to controlling the mechanical motion in optomechanics~\cite{Kippenberg2008rev,Aspelmeyer2012rev,Aspelmeyer2014} and nanomechanics~\cite{Blencowe2004,Poot2012}. However, it remains a challenge to generate macroscopically distinct superposition states~\cite{Nimmrichter2013} in mechanical resonators~\cite{Bose1999,Armour2002,Marshall2003, Tian2005, Cirac2011, Pepper2012, Tan2013, Akram2013, Ge2015, Liao2015}. Decoherence by quantum and thermal fluctuations can often destroy such superposition. Moreover, the natural mechanical displacement induced by a single photon in optomechanical systems is proportional to the ratio of the coupling rate to the mechanical frequency~\cite{Marshall2003}, $g_{0}/\omega_{M}$ [cf. Eq.~(\ref{Hamiltonian1})], which is of the order of $10^{-5}$ - $10^{-2}$ in realistic systems~\cite{Aspelmeyer2014}. To distinguish the single-photon mechanical displacement from its zero-point fluctuation, the ultrastrong coupling condition $g_{0}>\omega_{M}$ needs to be satisfied~\cite{Marshall2003}.

In this Letter, we propose an efficient approach for creating superposition of large-amplitude coherent states in a two-mode optomechanical system by introducing a sinusoidally modulated photon hopping between the two cavities. This modulated photon tunneling induces a near-resonant radiation-pressure force acting on the mechanical resonator, with an effective detuning much smaller than the original mechanical frequency, and hence increases the mechanical displacement generated by a single photon. One merit of this method is that the fidelities of the generated mechanical states are not affected by the decay of cavity photons. This feature enables the possibility to observe distinct mechanical superposition states in practical systems.

\emph{System.}---Consider a two-mode optomechanical system that consists of a free (left) cavity coupled to an optomechanical (right) cavity via a modulated photon-hopping interaction. The system is described by the Hamiltonian ($\hbar=1$)
\begin{eqnarray}
\hat{H}(t)&=&\omega_{c}(\hat{a}_{L}^{\dagger}\hat{a}_{L}+\hat{a}_{R}^{\dagger}\hat{a}_{R}) -\xi\omega_{0}\cos(\omega_{0}t)(\hat{a}_{L}^{\dagger}\hat{a}_{R}+\hat{a}_{R}^{\dagger}\hat{a}_{L})\nonumber \\
&&+\omega_{M}\hat{b}^{\dagger}\hat{b}-g_{0}\hat{a}_{R}^{\dagger}\hat{a}_{R}(\hat{b}+\hat{b}^{\dagger}),\label{Hamiltonian1}
\end{eqnarray}
where $\hat{a}_{L(R)}$ and $\hat{b}$ are the annihilation operators of the left (right) cavity mode and the mechanical mode, with resonant frequencies $\omega _{c}$ and $\omega_{M}$, respectively. The parameter $\omega_{0}$ is the modulation frequency, and $\xi$ is the dimensionless modulation amplitude of photon hopping between the two cavities. $g_{0}$ is the magnitude of the single-photon optomechanical coupling between the right cavity and the mechanical mode. Similar two-mode optomechanical systems have been proposed for studying quantum optics and quantum information missions~\cite{Miao2009,Dobrindt2010,Ludwig2012,Stannigel2012,Komar2013}.

In a rotating frame defined by the transformation operator $\hat{T}(t)=\hat{V}_{1}(t)\hat{V}_{2}(t)$ with $\hat{V}_{1}(t)=\exp\{-i[\omega_{c}(\hat{a}_{L}^{\dagger
}\hat{a}_{L}+\hat{a}_{R}^{\dagger}\hat{a}_{R})+\omega_{M}\hat{b}^{\dagger}\hat{b}]t\}$ and $\hat{V}_{2}(t)=\exp[i\xi\sin(\omega_{0}t)(\hat{a}_{L}^{\dagger}\hat{a}_{R}+\hat{a}_{R}^{\dagger}\hat{a}_{L})]$, and under the condition
$\vert \delta\vert,g_{0}/2\ll \omega_{0},\omega_{M}$,
we can obtain an effective Hamiltonian by the rotating-wave approximation (RWA) as~\cite{seeSM}
\begin{equation}
\hat{H}_{\textrm{RWA}}(t)=g(\hat{a}_{L}^{\dagger}\hat{a}_{L}-\hat{a}_{R}^{\dagger}\hat{a}_{R})
(\hat{b}e^{-i\delta t}+\hat{b}^{\dagger}e^{i\delta t}).\label{Happrox}
\end{equation}
Here, $g=g_{0}J_{2n_{0}}(2\xi)/2$ is the normalized coupling constant under a selected integer $n_{0}$ and $\delta=\omega_{M}-2n_{0}\omega_{0}$ is a modulation-induced detuning, where $J_{n}(z)$ is the Bessel function of the first kind, and $n_{0}$ corresponds to the near-resonance term in the Jacobi-Anger expansions of the sinusoidal factor in $\hat{V}_{2}(t)$.

The Hamiltonian Eq.~(\ref{Happrox}) describes a driven harmonic oscillator with an effective driving force $g\langle(\hat{a}_{L}^{\dagger}\hat{a}_{L}-\hat{a}_{R}^{\dagger}\hat{a}_{R})\rangle$ on a mechanical quadrature that rotates at a frequency $\delta$.
Under this form, the maximum mechanical displacement induced by a single photon is $2g/|\delta|$, which, by choosing proper $\xi$ and $\delta$, could be much larger than the displacement $2g_{0}/\omega_{M}$~\cite{Marshall2003} in the single-cavity case.
The resonance driving effect can be seen more clearly by introducing the symmetric and asymmetric modes of the two cavities~\cite{seeSM}: $\hat{a}_{\pm}=(\hat{a}_{L}\pm\hat{a}_{R})/\sqrt{2}$. In the representation of $\hat{a}_{\pm}$, the frequencies of modes $\hat{a}_{\pm}$ are modulated by periodic functions with frequency $\omega_{0}$, and hence the Floquet sideband modes (with frequencies $\omega_{c}+m\omega_{0}$ for integers $m$) will assist the transitions of the system. As a result, we can choose a proper $\omega_{0}$ such that the conditional displacement process becomes resonant or near resonant and other processes are far off resonant. The physical picture can also be understood in the time domain~\cite{seeSM}. By hopping a single photon into and out of the right cavity at the proper time, the mechanical effect of the single photon will be amplified because the displacement effect can be accumulated when the driving force and the mechanical oscillation are in phase. At the same time, modulation sidebands are designed to suppress other parametric processes and hence an enhanced radiation-pressure interaction can be obtained.

\emph{Generation of Yurke-Stoler-like states.}---To generate mechanical superposition states, we consider an initial state $\vert\psi(0)\rangle=\frac{1}{\sqrt{2}}(\vert 1\rangle_{L}\vert 0\rangle_{R}+\vert0\rangle_{L}\vert 1\rangle_{R})\vert 0\rangle_{M}$, where $|n=0,1\rangle_{L(R)}$ are cavity-field Fock states, and $\vert 0\rangle_{M}$ is the mechanical ground state prepared via ground state cooling~\cite{Teufel2011,Chan2011,Kippenberg2012}. Applying the propagator associated with $\hat{H}_{\textrm{RWA}}(t)$ on this initial state, followed by the transformation $\hat{T}(t)$, we derive the state
\begin{equation}
\vert\psi(t)\rangle=\frac{e^{i\vartheta}}{\sqrt{2}}[\vert 1\rangle_{L}\vert 0\rangle_{R}\vert\varphi_{L}(t)\rangle_{M}+\vert 0\rangle_{L}\vert 1\rangle_{R}\vert\varphi_{R}(t)\rangle_{M}],\label{staapproxopt}
\end{equation}
where $\vartheta(t)=-(\omega_{c}-g^{2}/\delta)t-(g^{2}/\delta^{2})\sin(\delta t)$ is a global phase factor. The two states
$\vert\varphi_{L}(t)\rangle_{M}=\cos(\mu/2)\vert \beta(t)\rangle_{M}+i\sin(\mu/2)\vert -\beta(t)\rangle_{M}$
and $\vert\varphi_{R}(t)\rangle_{M}=(\vert\varphi_{L}(t)\rangle_{M})|_{\beta\leftrightarrow-\beta}$ are Yurke-Stoler-like states~\cite{Yurke1986}, which are quantum superposition of coherent states $\vert \pm\beta(t)\rangle_{M}$, where $\beta(t)=[-2ig\sin(\delta t/2)/\delta]e^{-i(\omega_{M}-\delta/2)t}$ and $\mu(t)=2\xi\sin(\omega_{0}t)$. For the resonant case $\delta=0$, we have $\beta_{\textrm{res}}(t)=-igt\exp(-i\omega_{M}t)$. Equation~(\ref{staapproxopt}) describes a three-mode entangled state that involves two cavity modes and a mechanical mode. To generate mechanical superposition states $\vert\varphi_{L(R)}(t)\rangle_{M}$, we need to measure the states of the cavity field.
\begin{figure}[tbp]
\center
\includegraphics[bb=13 2 424 150, width=0.45 \textwidth]{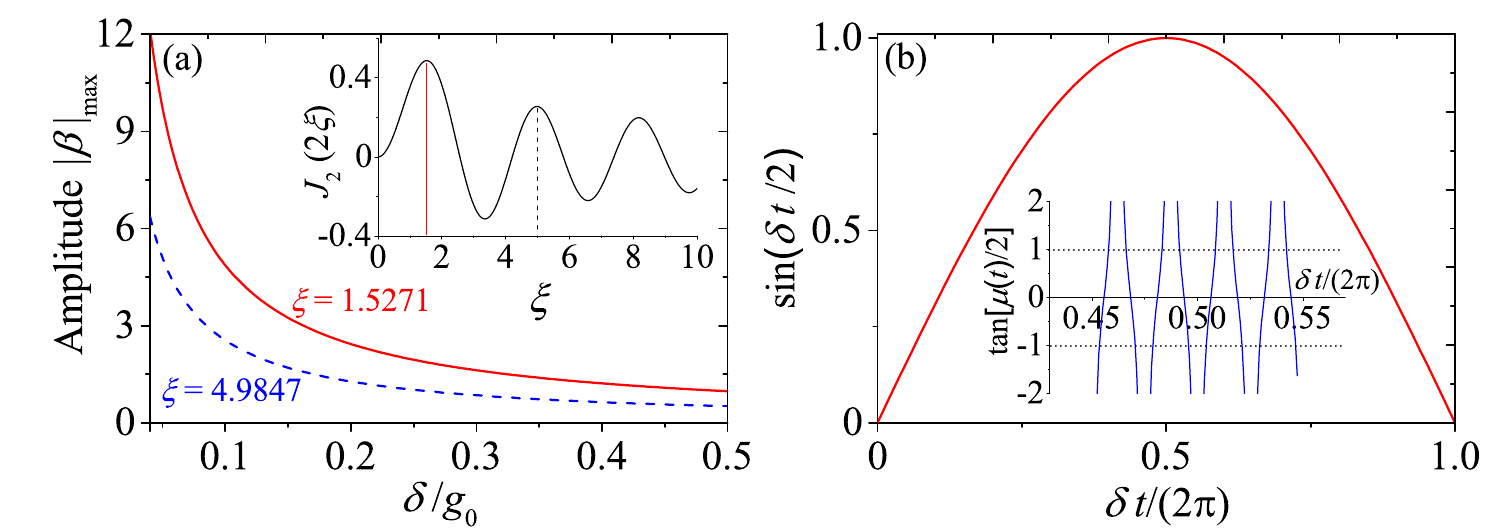}
\caption{(Color online) (a) The maximum amplitude $|\beta|_{\textrm{max}}$ versus $\delta/g_{0}$ at $\xi=1.5271$ and $4.9847$, which correspond to the peak values of the Bessel function $J_{2}(2\xi)$, as shown in the inset. (b) Time dependence of $\sin(\delta t/2)$ and $\tan[\mu(t)/2]$ near the positions that give the large oscillation amplitude and the equal probability superposition in states $|\varphi_{L(R)}(t)\rangle_{M}$, where $\omega_{M}/\delta=80$, $n_{0}=1$, and $\xi=1.527$.}
\label{Fig1}
\end{figure}

The maximum coherent amplitude, $|\beta|_{\textrm{max}}=2g/\delta$, is controllable by tuning the two parameters $\xi$ and $\omega_{0}$ based on the relations $g=g_{0}J_{2n_{0}}(2\xi)/2$ and $\delta=\omega_{M}-2n_{0}\omega_{0}$. We choose proper $n_{0}$ and optimal $\xi$ to reach peak values of the Bessel function $J_{2n_{0}}(2\xi)$, and tune the modulation frequency $\omega_{0}$ such that the value of $\delta$ can be changed continuously. In Fig.~\ref{Fig1}(a), we plot $|\beta|_{\textrm{max}}$ as a function of $\delta$ when the first two peak values of $J_{2}(2\xi)$ (with $n_{0}=1$) are taken (inset). A macroscopically distinct coherent amplitude can always be obtained by choosing $\delta<2g$ such that $|\beta|_{\textrm{max}}>1$ and then $|\langle-\beta|\beta\rangle|\ll 1$. In this case, the two coherent states become approximately distinguishable in phase space by proper quadrature measurements~\cite{Yurke1986,Bjork2004}.

The amplitude $|\beta|=(2g/\delta)|\sin(\delta t/2)|$ reaches its maximum values at times $t_{m}=(2m+1)\pi/\delta$ [i.e., $\sin(\delta t_{m}/2)=\pm1$] for non-negative integers $m$. Meanwhile, the relative probability amplitudes of the states $\vert\varphi_{L(R)}(t)\rangle_{M}$ depend on the time through $\mu(t)$. To observe strong evidence of quantum interference, one expects that the two components $\vert \pm\beta(t)\rangle_{M}$ appear with comparable probabilities. This leads to $\mu(\tau_{n})=2\xi \sin(\omega_{0}\tau_{n})\approx(n+1/2)\pi$ [i.e., $\tan[\mu(\tau_{n})/2]\approx\pm1$] for non-negative integers $n$. Near a given value of $t_{m}$, there are many $\tau_{n}$ satisfying the probability requirement because of $\omega_{0}\gg \delta$. Hence, we can choose proper time windows $\tau_{n}$ such that $|\beta(\tau_{n})|>1$. In Fig.~\ref{Fig1}(b), we plot the function $\sin(\delta t/2)$ and show the function $\tan[\mu(t)/2]$ around the time $t_{0}=\pi/\delta$ (inset). We can see that around $t_{0}$, there are many values of time satisfying the two requirements at the same time. In addition, the timing period of the measurement is slower than the periodic oscillation of the mechanical mode because of $\omega_0\approx\omega_M/2$. In realistic experiments, one can turn off the photon hopping at the detection time $t_{d}$ (the photon detection time, one of $\tau_{n}$ around $t_{0}$), then the evolution of the system can be approximated as a free evolution because the bare optomechanical coupling strength $g_{0}$ is much smaller than $\omega_{M}$. As a result, a wider time window can be obtained for implementing proper measurements for the cavities and the mechanical mode.

The above analyses show a trade-off between the displacement amplitude $|\beta|_{\textrm{max}}=2g/\delta$ and the state generation time $t_{0}=\pi/\delta$. We pursue a large $|\beta|_{\textrm{max}}$ for macroscopic superposition and a small $t_{0}$ for reducing the impact of the dissipations. In realistic simulations, we should choose a proper $\delta$ such that $|\beta|_{\textrm{max}}$ satisfies the requirement of macroscopicity and $t_{0}$ is as small as possible. It is also worth mentioning that the detection time can be shortened by utilizing the upslope rather than the peak of the amplitude function $|\sin(\delta t/2)|$ with a smaller $\delta$. For example, to obtain a displacement of $|\beta|_{\textrm{max}}=2$, the time for the resonant case $\delta=0$ is $t_{\textrm{res}}=2/g$, which is shorter than $t_{0}=\pi/g$ for the case $\delta=g$~\cite{seeSM}.

\emph{Effects of dissipations.}---To study the environmental fluctuation effects on the state generation scheme, we numerically simulate the state generation in the open system case, in which the evolution of our system is governed by the quantum master equation~\cite{seeSM}:
\begin{eqnarray}
\dot{\hat{\rho}}&=&i[\hat{\rho},\hat{H}(t)]+\gamma_{c}\mathcal{D}[\hat{a}_{L}]\hat{\rho}+\gamma_{c}\mathcal{D}[\hat{a}_{R}]\hat{\rho}\nonumber\\
&&+\gamma_{M}(n_{\textrm{th}}+1)\mathcal{D}[\hat{b}]\hat{\rho}+\gamma_{M}n_{\textrm{th}}\mathcal{D}[\hat{b}^{\dag}]\hat{\rho},
\end{eqnarray}
where $\mathcal{D}[\hat{o}]\hat{\rho}=\hat{o}\hat{\rho} \hat{o}^{\dagger}-(\hat{o}^{\dagger}\hat{o}\hat{\rho}+\hat{\rho} \hat{o}^{\dagger}\hat{o})/2$ is the standard Lindblad superoperator for photon and phonon dampings, $\gamma_{c}$ and $\gamma_{M}$ are the damping rates of the cavity fields and the mechanical mode, respectively, and $n_{\textrm{th}}$ is the thermal phonon occupation number. We numerically solve the master equation and calculate the reduced density matrix $\hat{\rho}^{(L)}_{M}(t)$ [$\hat{\rho}^{(R)}_{M}(t)$] of the mechanical mode~\cite{seeSM}, the probability $P_{L(R)}(t)$ of the photon in the left (right) cavity, and the fidelity $F_{s=L(R)}(t)=\,_{M}\!\langle\varphi_{s}(t)|\hat{\rho}^{(s)}_{M}(t)|\varphi_{s}(t)\rangle_{M}$ between the generated mechanical states and the target states.
\begin{figure}[tbp]
\center
\includegraphics[bb=2 0 415 329, width=0.45 \textwidth]{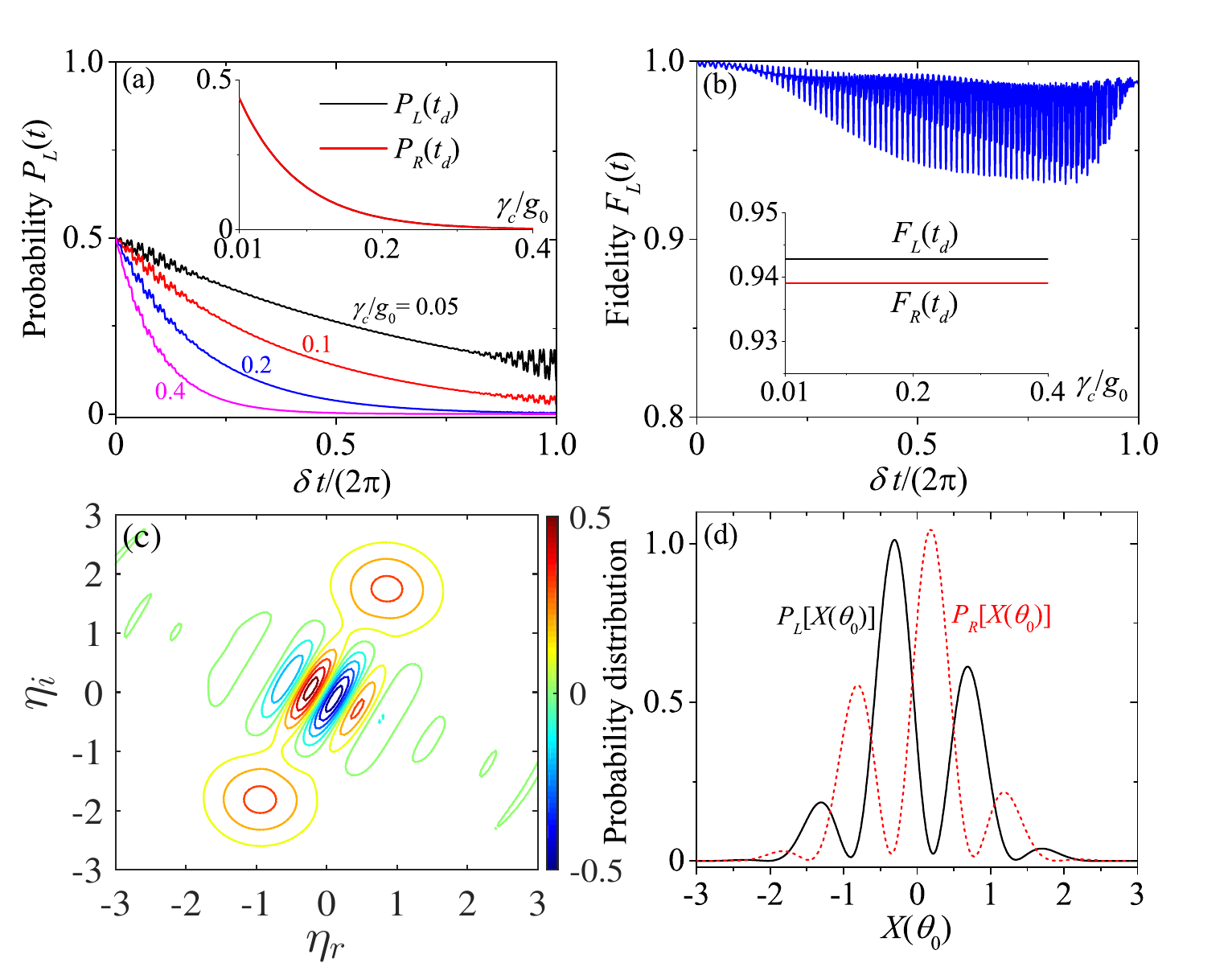}
\caption{(Color online) (a) The probability $P_{L}(t)$ and (b) the fidelity $F_{L}(t)$ versus $\delta t$ at selected values of the cavity-field decay rate $\gamma_{c}$. Insets: The probability $P_{L(R)}(t_{d})$ and the fidelity $F_{L(R)}(t_{d})$ at time $t_{d}$ versus $\gamma_{c}/g_{0}$. (c) The Wigner function $W_{L}(\eta)$ (with $\eta=\eta_{r}+i\eta_{i}$) and (d) the probability distribution of the rotated quadrature operator $P_{L}[X(\theta_{0})]$ for the state $\hat{\rho}^{(L)}_{M}(t_{d})$. Other parameters are $\omega_{M}/g_{0}=20$, $n_{0}=1$, $\xi=1.5271$, $\delta=g$, $\gamma_{M}/g_{0}=0.0001$, and $n_{\textrm{th}}=4$.}
\label{Fig2}
\end{figure}

In Fig.~\ref{Fig2}(a), we show the time dependence of the probability $P_{L}(t)$ at selected values of the cavity-field decay rate $\gamma_{c}$. Note that $P_{R}(t)$ has a similar pattern to $P_{L}(t)$ except for a slight oscillation [hereafter, we display only $P_{L}(t)$ and $F_{L}(t)$ for concision]. We see that $P_{L}(t)$ has an approximate exponential decay envelope with the corresponding $\gamma_{c}$ and slight oscillations. We also show the probabilities $P_{L}(t_{d})$ and $P_{R}(t_{d})$ at time $t_{d}$ as a function of $\gamma_{c}$ (inset). The curves indicate that $P_{L(R)}(t_{d})$ decreases with the increase of $\gamma_{c}$. About the fidelity, our numerical results show that the fidelities $F_{L}(t)$ and $F_{R}(t)$ have a similar pattern, and that the fidelities are independent of the decay rate $\gamma_{c}$, as shown in both the dynamics [Fig.~\ref{Fig2}(b)] and the fidelity at time $t_{d}$ (inset). Here the negligible difference between $F_{L}(t_{d})=0.943$ and $F_{R}(t_{d})=0.939$ is caused by the RWA, and it will disappear gradually with the increase of $\omega_{M}/g_{0}$. In the presence of photon dissipation, the photon could leak out of the cavities before the measurement. However, the normalization of the density matrices after the measurement ensures that in the selected experiments the photon remains in the cavities. Hence, the fidelity of the generated state is not affected by photon decay.

To evaluate the quantum coherence and interference effects in the generated superposition states $\hat{\rho}^{(L)}_{M}(t_{d})$ and  $\hat{\rho}^{(R)}_{M}(t_{d})$, we examine the Wigner function $W_{s=L(R)}(\eta)=\frac{2}{\pi}\textrm{Tr}[\hat{D}^{\dag}(\eta)\hat{\rho}^{(s)}_{M}(t_{d})\hat{D}(\eta)(-1)^{\hat{b}^{\dag}\hat{b}}]$~\cite{Barnettbook}, where $\hat{D}(\eta)=\exp(\eta \hat{b}^{\dag}-\eta^{\ast}\hat{b})$ is the displacement operator.
It can be seen from the relation $\vert\varphi_{R}(t_{d})\rangle_{M}=(\vert\varphi_{L}(t_{d})\rangle_{M})|_{\beta\leftrightarrow -\beta}$ that the Wigner function $W_{R}(\eta)$ should be a rotation of $W_{L}(\eta)$ by $\pi$ about the origin in phase space. We perform the simulations with the parameters in Fig.~\ref{Fig2} and find that there is also a negligible difference between $W_{R}(\eta)$ and the $\pi$-rotated $W_{L}(\eta)$. The difference disappears gradually with the increase of $\omega_{M}/g_{0}$. We also find that the Wigner functions are independent of the cavity-field decay rate, in accordance with the fidelities. In Fig.~\ref{Fig2}(c) we display the Wigner function $W_{L}(\eta)$ of the state $\hat{\rho}^{(L)}_{M}(t_{d})$. We see obvious interference evidence in this Wigner function.

The quantum superposition properties can also be seen in the probability distribution $P_{s=L(R)}[X(\theta)]=\,_{M}\!\langle X(\theta)|\hat{\rho}^{(s)}_{M}(t_{d})|X(\theta)\rangle_{M}$ of the rotated quadrature operator $\hat{X}(\theta)=(\hat{b} e^{-i\theta}+\hat{b}^{\dagger} e^{i\theta})/\sqrt{2}$~\cite{Milburnbook}, where $|X(\theta)\rangle_{M}$ is the eigenstate of $\hat{X}(\theta)$: $\hat{X}(\theta)|X(\theta)\rangle_{M}=X(\theta)|X(\theta)\rangle_{M}$.
In Fig.~\ref{Fig2}(d), we plot the probability distributions $P_{L}[X(\theta_{0})]$ and $P_{R}[X(\theta_{0})]$ as functions of $X(\theta_{0})$. Here the rotation angle is chosen as $\theta_{0}=\arg[\beta(t_{d})]-\pi/2$, which means that the quadrature direction is perpendicular to the link line between the locations of the two superposed coherent amplitudes. The interference is maximum in this direction because the two coherent states are projected onto the quadrature such that they overlap exactly. The oscillation in the curves is a distinct evidence of the quantum interference between the superposition components. We notice that $P_{L}[X(\theta_{0})]$ and $P_{R}[X(\theta_{0})]$ are approximately symmetric to each other about the vertical axis $X(\theta_{0})=0$, in accordance with the negligible difference between $F_{L}(t)$ and $F_{R}(t)$.
\begin{figure}[tbp]
\center
\includegraphics[bb=2 0 415 329, width=0.45 \textwidth]{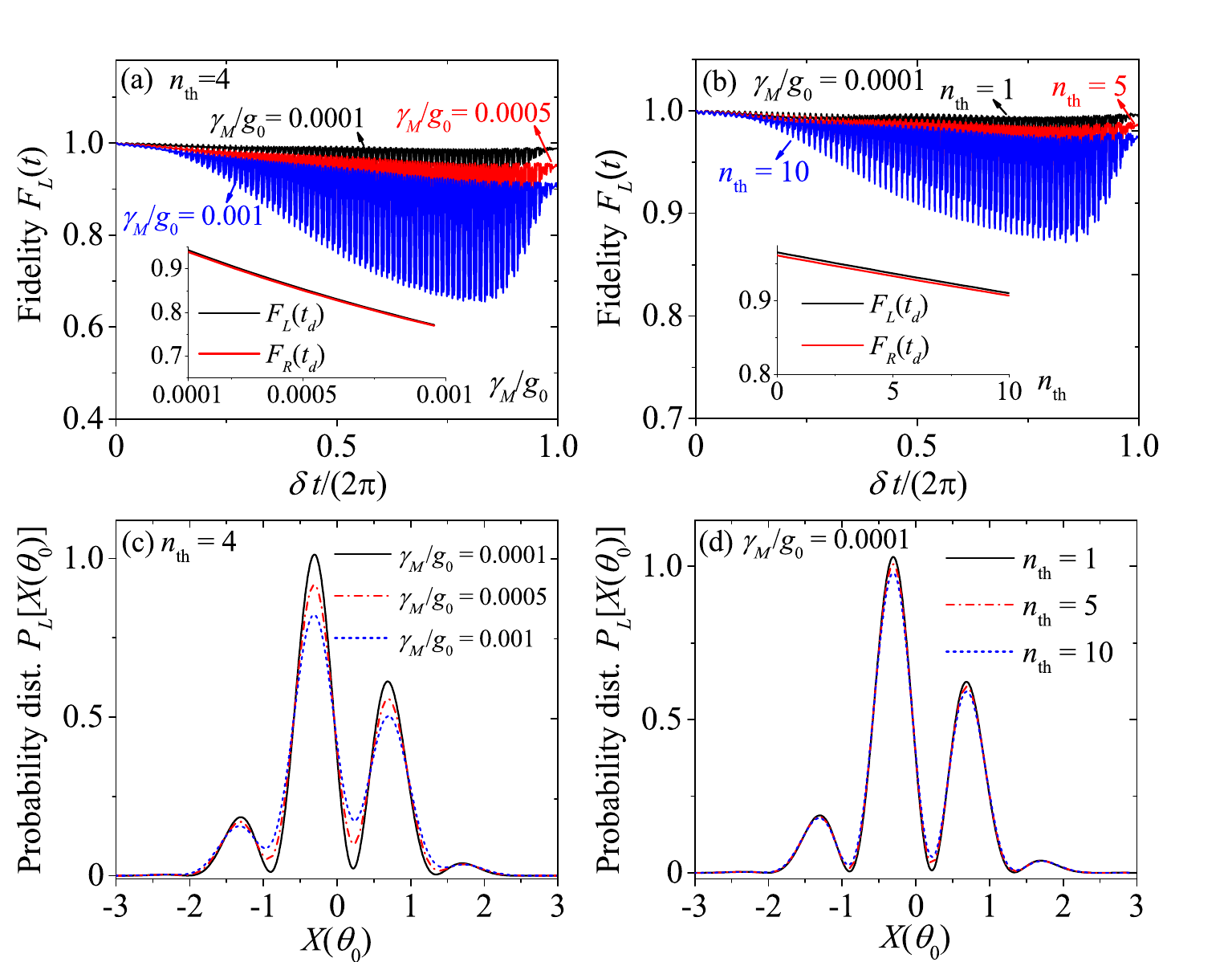}
\caption{(Color online) The fidelity $F_{L}(t)$ versus $\delta t$ at selected values of (a) the mechanical decay rate $\gamma_{M}/g_{0}$ and (b) the thermal phonon occupation number $n_{\textrm{th}}$. Insets: the fidelity $F_{L(R)}(t_{d})$ at time $t_{d}$ versus $\gamma_{M}/g_{0}$ and $n_{\textrm{th}}$. The probability distribution of the rotated quadrature operator $P_{L}[X(\theta_{0})]$ of the state $\hat{\rho}^{(L)}_{M}(t_{d})$ at selected values of (c) $\gamma_{M}/g_{0}$ and (d) $n_{\textrm{th}}$. Other parameters are $\omega_{M}/g_{0}=20$, $n_{0}=1$, $\xi=1.5271$, $\delta=g$, and $\gamma_{c}/g_{0}=0.2$.}
\label{Fig3}
\end{figure}

We also investigate the influence of mechanical noise on the state generation. Our numerical simulations verify that the probabilities $P_{L}(t)$ and $P_{R}(t)$ are independent of $\gamma_{M}$ and $n_{\textrm{th}}$. On the contrary, the mechanical dissipations affect the fidelities $F_{L}(t)$ and $F_{R}(t)$. In Figs.~\ref{Fig3}(a) and~\ref{Fig3}(b), we show the dynamics of the fidelity $F_{L}(t)$ at selected values of $\gamma_{M}$ and $n_{\textrm{th}}$, respectively. We can see $F_{L}(t)$ becomes worse for larger values of $\gamma_{M}$ and $n_{\textrm{th}}$. In addition, the fidelities $F_{L}(t_{d})$ and $F_{R}(t_{d})$ at time $t_{d}$ decrease with the increase of $\gamma_{M}$ and $n_{\textrm{th}}$ (insets). In Figs.~\ref{Fig3}(c) and~\ref{Fig3}(d), we plot the probability distribution $P_{L}[X(\theta_{0})]$ for the state $\hat{\rho}^{(L)}_{M}(t_{d})$ with the parameters in Figs.~\ref{Fig3}(a) and~\ref{Fig3}(b), respectively [we can know $P_{R}[X(\theta_{0})]$ from the approximate symmetry between $P_{L}[X(\theta_{0})]$ and $P_{R}[X(\theta_{0})]$]. Here we can see that the oscillatory feature of $P_{L}[X(\theta_{0})]$ disappears gradually with the increase of $\gamma_{M}$ and $n_{\textrm{th}}$. These results imply that mechanical dissipations will destroy the quantum coherence and interference effects in the generated mechanical superposition. In our simulations, quantum interference evidence can be seen, and good fidelities ($>0.9$) can be obtained.

\emph{Discussions.}---Our state generation approach is general and it can be principally implemented in various optomechanical setups. Below, we focus our discussion on electromechanical systems with cavities in the microwave regime. For such systems, the photon hopping between superconducting resonators can be realized via Josephson junction coupling~\cite{Devoret1997}. The initial Bell state of the cavity fields can be prepared by a superconducting qubit, as realized in circuit-QED systems~\cite{Hofheinz2008,Hofheinz2009}. In particular, a nonperfect photon loading (i.e., containing the zero-photon component) does not affect the fidelity but decreases the success probability of the generated mechanical states because all the couplings will be frozen when there are no photons in the two cavities. The photon states in the superconducting resonators can be measured via superconducting quits~\cite{Schuster2007}. In addition, the generated mechanical superposition states can be measured by the technique of quantum state reconstruction~\cite{Vitali2007,Lu2015,Vanner2015}. We use another cavity mode (in the same resonator) to build a connection between the mechanical mode and the output field. By detecting the quadrature of the output field, we can obtain the information of the mechanical states~\cite{seeSM}.

The parameter conditions for implementation of this scheme are: $g_{0}\ll \omega_{M}$, the ratio $g_{0}/\gamma_{c}$ should be moderately larger than $1$ for a high success probability (for example, $g_{0}/\gamma_{c}=5$ - $10$ corresponds to the success probability $0.08$ - $0.285$), and $n_{\textrm{th}}\ll g_{0}/(4\pi\gamma_{M})$. Below, we analyze the conditions in detail~\cite{seeSM}. (i) For state generation purposes, we choose $\delta<2g\ll \omega_{M}$, then the RWA condition can be simplified as $g_{0}\ll \omega_{M}$, which is consistence with the current experimental situation~\cite{Aspelmeyer2014}: $g_{0}/\omega_{M}$ is of the order of $10^{-5}$ - $10^{-3}$. (ii) The photon decay does not affect the fidelity, but it affects the probability by $\mathcal{P}\approx e^{-4\pi\gamma_{c}/g_{0}}$ at $\delta=g$. Currently, the value of $g_{0}/\gamma_{c}$ is $10^{-4}$ - $10^{-2}$~\cite{Aspelmeyer2012rev}. This value can be increased by either increasing $g_{0}$ or decreasing $\gamma_{c}$. In electromechanical systems, $\gamma_{c}=2\pi\times170$ kHz~\cite{Teufel2011B} and $\gamma_{c}=2\pi\times118$ kHz~\cite{Toth2016} have been reported. The value of $\gamma_{c}$ can be further decreased to be dozens of kilohertz~\cite{Megrant2012}. The largest value of $g_{0}$ reported in electromechanics is $2\pi\times 460$ Hz~\cite{Teufel2011B}, and theoretic estimations indicate that it can reach megahertz by utilizing the nonlinearity in Josephson junction~\cite{Rimberg2014,Heikkila2014}. Therefore, $g_{0}/\gamma_{c}>5$ should be accessible in the near future. In particular, in the resonant case $\delta=0$ and at $|\beta|_{\textrm{max}}=1$, the success probability can be improved to be $\mathcal{P}\approx e^{-4\gamma_{c}/g_{0}}$, which takes $\mathcal{P}=0.14$ - $0.45$ for $g_{0}/\gamma_{c}=2$ - $5$. (iii) The thermal phonon number $n_{\textrm{th}}$ should be small such that the state generation time is much shorter than the characteristic coherence time of the phonons, i.e., $t_{d}\approx4\pi/g_{0}\ll 1/(\gamma_{M}n_{\textrm{th}})$, which leads to $n_{\textrm{th}}\ll g_{0}/(4\pi\gamma_{M})$. Currently, the ratio $g_{0}/\gamma_{M}$ is $10^{1}$ - $10^{2}$~\cite{Aspelmeyer2014} (this value can be increased to $10^{4}$ when $g_{0}$ is increased as described above). In a low-temperature environment, $n_{\textrm{th}}<30$ can be obtained. For example, at $T=10$ mK~\cite{Toth2016}, we have $n_{\textrm{th}}\approx20$ at $\omega_{M}=2\pi\times10$ MHz. Therefore the condition $n_{\textrm{th}}\ll g_{0}/(4\pi\gamma_{M})$ can be satisfied in electromechanics. Based on the above discussions, we suggest the parameters to be $\omega_{c}=2\pi\times$ ($5$ - $10$) GHz, $\gamma_{c}=2\pi\times$ ($25$ - $200$) kHz, $\omega_{M}=2\pi\times10$ MHz, $\gamma_{M}=2\pi\times$ ($50$ - $500$) Hz, and $g_{0}=2\pi\times500$ kHz, which are consistent with the values used in our simulations~\cite{seeSM}.

\emph{Conclusions.}---We have proposed an efficient method for creating macroscopically distinct superposition states in a mechanical resonator. This method is based on the introduction of a modulated photon-hopping interaction in a two-mode optomechanical system  to produce large effective single-photon optomechanical coupling. Numerical simulations demonstrate that our method works well in the presence of dissipations, and can be realized in a wide parameter range.

\emph{Acknowledgments.}---The authors are supported by the DARPA ORCHID program through AFOSR and
the National Science Foundation under Award No. NSF-DMR-0956064.

\newpage
\onecolumngrid
\newpage
\begin{center}
\textbf{\large Supplementary materials for ``Macroscopic Quantum Superposition in Cavity Optomechanics"}
\end{center}
\setcounter{equation}{0}
\setcounter{figure}{0}
\setcounter{table}{0}
\setcounter{page}{1}
\makeatletter
\renewcommand{\theequation}{S\arabic{equation}}
\renewcommand{\thefigure}{S\arabic{figure}}

This document consists of four parts: (I) Derivation of the approximate Hamiltonian $\hat{H}_{\textrm{RWA}}(t)$ and the physical picture of mechanical displacement enhancement; (II) Detailed calculations in the closed-system case; (III) Detailed calculations in the open-system case; (IV) Discussions on the experimental implementation.

\section{I. Derivation of the approximate Hamiltonian $\hat{H}_{\textrm{RWA}}(t)$ and the physical picture of mechanical displacement enhancement}

\subsection{A. Derivation of the approximate Hamiltonian $\hat{H}_{\textrm{RWA}}(t)$}

In this section, we present a detailed derivation of the approximate Hamiltonian $\hat{H}_{\textrm{RWA}}(t)$ given in the main text.
We start from the full Hamiltonian of the system ($\hbar=1$)
\begin{equation}
\hat{H}(t)=\omega_{c}(\hat{a}_{L}^{\dagger}\hat{a}_{L}+\hat{a}_{R}^{\dagger}\hat{a}_{R})
-\xi\omega_{0}\cos(\omega_{0}t)(\hat{a}_{L}^{\dagger}\hat{a}_{R}+\hat{a}_{R}^{\dagger}\hat{a}_{L})
+\omega_{M}\hat{b}^{\dagger}\hat{b}-g_{0}\hat{a}_{R}^{\dagger}\hat{a}_{R}(\hat{b}+\hat{b}^{\dagger}),\label{fullHamilt}
\end{equation}
where $\hat{a}_{L(R)}$ and $\hat{b}$ are the annihilation operators of the left (right)
cavity field and the mechanical mode, with the resonance
frequencies $\omega_{c}$ ($\omega_{c}$) and $\omega_{M}$, respectively. The parameters $\omega_{0}$ is the modulation frequency in the photon-hopping interaction between the two cavities, and $\xi$ is a dimensionless coupling constant. The parameter $g_{0}$ is the single-photon optomechanical-coupling strength between the right cavity field and the mechanical mode.

To see clearly the action of the photon-hopping modulation, we perform a transformation defined by $\hat{T}(t)=\hat{V}_{1}(t)\hat{V}_{2}(t)$ to the Hamiltonian $\hat{H}(t)$, where
\begin{equation}
\hat{V}_{1}(t)=\exp\left\{-i[\omega_{c}(\hat{a}_{L}^{\dagger
}\hat{a}_{L}+\hat{a}_{R}^{\dagger}\hat{a}_{R})+\omega_{M}\hat{b}^{\dagger}\hat{b}]t\right\},\hspace{1 cm}
\hat{V}_{2}(t)=\exp\left[i\xi\sin(\omega_{0}t)(\hat{a}_{L}^{\dagger}\hat{a}_{R}+\hat{a}_{R}^{\dagger}\hat{a}_{L})\right].
\end{equation}
Then the transformed Hamiltonian becomes
\begin{eqnarray}
\hat{\tilde{H}}(t)&=&\hat{T}^{\dagger}(t) \hat{H}(t)\hat{T}(t)-i\hat{T}^{\dagger}(t)\dot{\hat{T}}(t)\nonumber\\
&=&\hat{V}^{\dag}_{2}(t)\hat{V}^{\dag}_{1}(t)\hat{H}(t)\hat{V}_{1}(t)\hat{V}_{2}(t)-i\hat{V}^{\dag}_{2}(t)\hat{V}^{\dag}_{1}(t)
[\dot{\hat{V}}_{1}(t)\hat{V}_{2}(t)+\hat{V}_{1}(t)\dot{\hat{V}}_{2}(t)]\nonumber\\
&=&\hat{V}^{\dag}_{2}(t)[\hat{V}^{\dag}_{1}(t)\hat{H}(t)\hat{V}_{1}(t)-i\hat{V}^{\dag}_{1}(t)\dot{\hat{V}}_{1}(t)]\hat{V}_{2}(t)
-i\hat{V}^{\dag}_{2}(t)\dot{\hat{V}}_{2}(t)\nonumber\\
&\equiv&\hat{V}_{2}^{\dagger }(t) \hat{H}_{1}(t)
\hat{V}_{2}(t) -i\hat{V}_{2}^{\dagger }(t) \dot{\hat{V}}_{2}(t),
\end{eqnarray}
which can be obtained by two steps of transformations. Here the first-step transformation is
\begin{eqnarray}
\hat{H}_{1}(t)&=&\hat{V}_{1}^{\dagger}(t)\hat{H}(t)\hat{V}_{1}(t)-i\hat{V}_{1}^{\dagger}(t)\dot{\hat{V}}_{1}(t)\notag \\
&=&-\xi\omega_{0}\cos(\omega_{0}t)(\hat{a}_{L}^{\dagger}\hat{a}_{R}+\hat{a}_{R}^{\dagger}\hat{a}_{L})-g_{0}\hat{a}_{R}^{\dagger }\hat{a}_{R}(\hat{b}e^{-i\omega_{M}t}+\hat{b}^{\dagger }e^{i\omega_{M}t}),
\end{eqnarray}
and the second-step transformation is
\begin{eqnarray}
\hat{\tilde{H}}(t)&=&\hat{V}_{2}^{\dagger }(t) \hat{H}_{1}(t)
\hat{V}_{2}(t) -i\hat{V}_{2}^{\dagger }(t) \dot{\hat{V}}_{2}(t)\notag \\
&=&-\frac{g_{0}}{2}\left[(\hat{a}_{L}^{\dagger}\hat{a}_{L}+\hat{a}_{R}^{\dagger}\hat{a}_{R})-\cos[2\xi\sin(\omega_{0}t)](\hat{a}_{L}^{\dagger }\hat{a}_{L}-\hat{a}_{R}^{\dagger}\hat{a}_{R})+i\sin[2\xi\sin(\omega_{0}t)](\hat{a}_{R}^{\dagger}\hat{a}_{L}-\hat{a}_{L}^{\dagger}\hat{a}_{R})\right]\nonumber\\
&&\times(\hat{b}e^{-i\omega_{M}t}+\hat{b}^{\dag}e^{i\omega_{M}t}),\label{Htildeopt}
\end{eqnarray}
where we used the relations
\begin{subequations}
\begin{align}
\hat{V}_{2}^{\dagger}(t)\hat{a}_{R}\hat{V}_{2}(t)&=\cos[\xi\sin(\omega_{0}t)]\hat{a}_{R}+i\sin[\xi\sin(\omega_{0}t)]\hat{a}_{L},\\
\hat{V}_{2}^{\dagger}(t)\hat{a}^{\dag}_{R}\hat{V}_{2}(t)&=\cos[\xi\sin(\omega_{0}t)]\hat{a}^{\dag}_{R}-i\sin[\xi\sin(\omega_{0}t)]\hat{a}^{\dag}_{L}.
\end{align}
\end{subequations}

Using the Jacobi-Anger expansions
\begin{subequations}
\label{JAexapnsions}
\begin{align}
\cos[2\xi\sin(\omega_{0}t)]&=J_{0}(2\xi)+2\sum_{n=1}^{\infty }J_{2n}(2\xi)\cos(2n\omega_{0}t),\\
\sin[2\xi\sin(\omega_{0}t)]&=2\sum_{n=1}^{\infty}J_{2n-1}(2\xi)\sin[(2n-1)\omega_{0}t],
\end{align}
\end{subequations}
with $J_{n}(z)$ being the Bessel function of the first kind, the Hamiltonian $\hat{\tilde{H}}(t)$ can be expanded into a summation of many oscillating terms.
By observing Eqs.~\eqref{Htildeopt} and (\ref{JAexapnsions}), we know that the oscillation frequencies of these terms are $\omega_{M}\pm n\omega_{0}$ for natural numbers $n$. Therefore, when the frequency step $\omega_{0}$ is much larger than $g_{0}/2$, we can pick up a resonant or near-resonant term by choosing proper $\omega_{0}$ and $n$. For example, we choose a characteristic number $n_{0}$ ($n_{0}>0$), and assume that the term with a detuning $\delta =(\omega_{M}-2n_{0}\omega_{0})$ is the near-resonant term. Then we have $\omega_{M}=\delta+2n_{0}\omega_{0}$ and the Hamiltonian $\hat{\tilde{H}}(t)$ can be expressed as
\begin{eqnarray}
\hat{\tilde{H}}(t)&=&-\frac{g_{0}}{2}\left\{[1-J_{0}(2\xi)]\hat{a}_{L}^{\dagger }\hat{a}_{L}+[1+J_{0}(2\xi)]
\hat{a}_{R}^{\dagger}\hat{a}_{R}\right\}(\hat{b}e^{-i\omega_{M}t}+\textrm{H.c.})\nonumber\\
&&+\frac{g_{0}}{2}(\hat{a}_{L}^{\dagger}\hat{a}_{L}-\hat{a}_{R}^{\dagger}\hat{a}_{R})\left\{J_{2}(2\xi)[
(\hat{b}e^{-i[\delta+2(n_{0}-1)\omega_{0}]t}+\textrm{H.c.})
+(\hat{b}e^{-i[\delta+2(n_{0}+1)\omega_{0}]t}+\textrm{H.c.})]\right.\nonumber\\
&&\left.+J_{4}(2\xi)[
(\hat{b}e^{-i[\delta+2(n_{0}-2)\omega_{0}]t}+\textrm{H.c.})
+(\hat{b}e^{-i[\delta+2(n_{0}+2)\omega_{0}]t}+\textrm{H.c.})]+\cdots\right.\nonumber\\
&&\left.+J_{2n_{0}}(2\xi)[\underline{(\hat{b}e^{-i\delta t}+\textrm{H.c.})}
+(\hat{b}e^{-i[\delta+4n_{0}\omega_{0}]t}+\textrm{H.c.})]+\cdots\right.\nonumber\\
&&\left.+J_{2n}(2\xi)[(\hat{b}e^{-i[\delta+2(n_{0}-n)\omega_{0}]t}+\textrm{H.c.})
+(\hat{b}e^{-i[\delta+2(n_{0}+n)\omega_{0}]t}+\textrm{H.c.})]+\cdots\right\}\nonumber\\
&&+\frac{g_{0}}{2}(\hat{a}_{L}^{\dagger}\hat{a}_{R}-\hat{a}_{R}^{\dagger}\hat{a}_{L})\left\{J_{1}(2\xi)[
(\hat{b}e^{-i[\delta+(2n_{0}-1)\omega_{0}]t}-\textrm{H.c.})
-(\hat{b}e^{-i[\delta+(2n_{0}+1)\omega_{0}]t}-\textrm{H.c.})]\right.\nonumber\\
&&\left.+J_{3}(2\xi)[(\hat{b}e^{-i[\delta+(2n_{0}-3)\omega_{0}]t}-\textrm{H.c.})
-(\hat{b}e^{-i[\delta+(2n_{0}+3)\omega_{0}]t}-\textrm{H.c.})]+\cdots\right.\nonumber\\
&&\left.+J_{2n_{0}-1}(2\xi)[(\hat{b}e^{-i[\delta+\omega_{0}]t}-\textrm{H.c.})
-(\hat{b}e^{-i[\delta+(4n_{0}-1)\omega_{0}]t}-\textrm{H.c.})]+\cdots\right.\nonumber\\
&&\left.+J_{2n-1}(2\xi)[(\hat{b}e^{-i\{\delta+[2n_{0}-(2n-1)]\omega_{0}\}t}-\textrm{H.c.})
-(\hat{b}e^{-i\{\delta+[2n_{0}+(2n-1)]\omega_{0}\}t}-\textrm{H.c.})]+\cdots\right\}.\label{Htransall}
\end{eqnarray}
We can see clearly from Eq.~(\ref{Htransall}) that the oscillation frequency of the underlined term is $\delta$, and the oscillation frequencies of all other terms differ from $\delta$ at $|m\omega_{0}|$ for nonzero integers $m$.
Under the condition
\begin{equation}
|\delta|,\; \frac{g_{0}}{2}\ll \omega_{0},\;\omega_{M},\label{SMRWAcond}
\end{equation}
we can neglect the high-frequency oscillating terms by the rotating-wave approximation (RWA) and obtain an approximate
Hamiltonian as
\begin{equation}
\hat{H}_{\text{RWA}}(t)= g(\hat{a}_{L}^{\dagger}\hat{a}_{L}-\hat{a}_{R}^{\dagger}\hat{a}_{R})(\hat{b}e^{-i\delta t}+\hat{b}^{\dagger}e^{i\delta t}),\label{Heff}
\end{equation}
where $g=g_{0}J_{2n_{0}}(2\xi)/2$. It should be noted that the detuning $|\delta|$ in Eq.~(\ref{SMRWAcond}) should not be much larger than $g$ so that the target term given in Eq.~(\ref{Heff}) is the dominating term in $\hat{\tilde{H}}(t)$. Further, for generation of mechanical superposition states with macroscopically distinguishable coherent-state components, the detuning $|\delta|$ should be chosen as $2g/|\delta|>1$ such that the coherent state components in the generated mechanical states can be distinguished from each other.

We can also understand the physical mechanism of resonance in the representation of
symmetric and antisymmetric cavity modes. For the two coupled cavities, the symmetric and antisymmetric cavity modes can be defined as
\begin{eqnarray}
\hat{a}_{+}=(\hat{a}_{L}+\hat{a}_{R})/\sqrt{2},\hspace{1 cm}
\hat{a}_{-}=(\hat{a}_{L}-\hat{a}_{R})/\sqrt{2}.
\end{eqnarray}
In the representation of $\hat{a}_{\pm}$, the Hamiltonian~(\ref{fullHamilt}) becomes
\begin{eqnarray}
\hat{H}(t)&=&[\omega_{c}-\xi\omega_{0}\cos(\omega_{0}t)]\hat{a}_{+}^{\dagger}\hat{a}_{+}
+[\omega_{c}+\xi\omega_{0}\cos(\omega_{0}t)]\hat{a}_{-}^{\dagger}\hat{a}_{-}+\omega_{M}\hat{b}^{\dagger}\hat{b}\notag\\
&&-\frac{g_{0}}{2}[(\hat{a}_{+}^{\dagger}\hat{a}_{+}+\hat{a}_{-}^{\dagger}\hat{a}_{-})-(\hat{a}_{+}^{\dagger}\hat{a}_{-}+
\hat{a}_{-}^{\dagger}\hat{a}_{+})](\hat{b}+\hat{b}^{\dagger}),\label{Hsymmrepre}
\end{eqnarray}
where we used the relations
\begin{eqnarray}
\hat{a}_{L}^{\dagger}\hat{a}_{L}+\hat{a}_{R}^{\dagger}\hat{a}_{R}&=&\hat{a}_{+}^{\dagger}\hat{a}_{+}+\hat{a}_{-}^{\dagger}\hat{a}_{-},\hspace{1 cm}
\hat{a}_{L}^{\dagger}\hat{a}_{L}-\hat{a}_{R}^{\dagger}\hat{a}_{R}=\hat{a}_{+}^{\dagger}\hat{a}_{-}+\hat{a}_{-}^{\dagger}\hat{a}_{+},\notag\\
\hat{a}_{L}^{\dagger}\hat{a}_{R}+\hat{a}_{R}^{\dagger}\hat{a}_{L}&=&\hat{a}_{+}^{\dagger}\hat{a}_{+}-\hat{a}_{-}^{\dagger}\hat{a}_{-},\hspace{1 cm}
\hat{a}_{L}^{\dagger }\hat{a}_{R}-\hat{a}_{R}^{\dagger }\hat{a}_{L}=-(\hat{a}_{+}^{\dagger}\hat{a}_{-}-\hat{a}_{-}^{\dagger }\hat{a}_{+}).
\label{relationsymbare}
\end{eqnarray}
As shown in Eq.~(\ref{Hsymmrepre}), the frequencies of the symmetric and antisymmetric cavity modes $\hat{a}_{\pm}$ are $\omega_{c}\mp\xi\omega_{0}\cos(\omega_{0}t)$, which are periodic functions with frequency $\omega_{0}$. According to the Floquet theory, the frequencies of these Floquet sidebands are $\omega_{c}+n\omega_{0}$ for integers $n$. These sidebands will be involved into the transitions of the photons and phonons, and hence we can design proper modulation frequency $\omega_{0}$ such that the desired physical processes to be resonant or near-resonant and other processes to be far-off-resonant. To see this point, we turn to a rotating frame defined by the transformation
\begin{eqnarray}
\hat{V}(t)&=&\exp \left(-i\{[\omega_{c}t-\xi\sin(\omega_{0}t)]\hat{a}_{+}^{\dagger}\hat{a}_{+}+[
\omega _{c}t+\xi\sin(\omega_{0}t)]\hat{a}_{-}^{\dagger }\hat{a}_{-}+\omega _{M}t\hat{b}^{\dagger }\hat{b}\}\right)
=\hat{T}(t).
\end{eqnarray}
In this frame, the transformed Hamiltonian becomes
\begin{eqnarray}
\hat{\tilde{H}}(t)&=&-\frac{g_{0}}{2}(\hat{a}_{+}^{\dagger }\hat{a}_{+}+\hat{a}_{-}^{\dagger}\hat{a}_{-})(
\hat{b}e^{-i\omega _{M}t}+\hat{b}^{\dagger }e^{i\omega _{M}t})  \notag\\
&&+\frac{g_{0}}{2}\left(\hat{a}_{+}^{\dagger}\hat{a}_{-}e^{-2i\xi\sin(\omega_{0}t)}
+\hat{a}_{-}^{\dagger}\hat{a}_{+}e^{2i\xi\sin(\omega_{0}t)}\right)(\hat{b}e^{-i\omega_{M}t}+\hat{b}^{\dagger}e^{i\omega_{M}t}).
\end{eqnarray}
By expanding the functions $\exp[\pm2i\xi\sin(\omega_{0}t)]$ using the Jacobi-Anger equality, the transformed Hamiltonian becomes
\begin{eqnarray}
\hat{\tilde{H}}(t)&=&-\frac{g_{0}}{2}(\hat{a}_{+}^{\dagger }\hat{a}_{+}+\hat{a}_{-}^{\dagger}\hat{a}_{-})(\hat{b}e^{-i\omega_{M}t}+\hat{b}^{\dagger}e^{i\omega_{M}t})
+\frac{g_{0}}{2}J_{0}(2\xi)(\hat{a}_{+}^{\dagger}\hat{a}_{-}+\hat{a}_{-}^{\dagger}\hat{a}_{+})
(\hat{b}e^{-i\omega_{M}t}+\hat{b}^{\dagger}e^{i\omega _{M}t})\nonumber\\
&&+\frac{g_{0}}{2}\sum\limits_{n=1}^{\infty }J_{2n}(2\xi)(\hat{a}_{+}^{\dagger}\hat{a}_{-}+\hat{a}_{-}^{\dagger }\hat{a}_{+})
[\hat{b}e^{-i(\omega_{M}+2n\omega_{0})t}+\hat{b}^{\dagger}e^{i(\omega_{M}+2n\omega_{0})t}]\nonumber\\
&&+\frac{g_{0}}{2}\sum\limits_{n=1}^{\infty }J_{2n-1}(2\xi)(\hat{a}_{+}^{\dagger}\hat{a}_{-}-\hat{a}_{-}^{\dagger}
\hat{a}_{+})[\hat{b}e^{-i[\omega_{M}+(2n-1)\omega_{0}]t}-\hat{b}^{\dagger}e^{i[\omega_{M}+(2n-1)\omega_{0}]t}]\nonumber\\
&&+\frac{g_{0}}{2}\sum\limits_{n=1}^{\infty }\underline{J_{2n}(2\xi)(\hat{a}_{+}^{\dagger}\hat{a}_{-}+\hat{a}_{-}^{\dagger}\hat{a}_{+})
[\hat{b}e^{-i(\omega_{M}-2n\omega_{0})t}+\hat{b}^{\dagger}e^{i(\omega_{M}-2n\omega_{0})t}]}\nonumber\\
&&-\frac{g_{0}}{2}\sum\limits_{n=1}^{\infty }J_{2n-1}(2\xi)(\hat{a}_{+}^{\dagger}\hat{a}_{-}-\hat{a}_{-}^{\dagger}
\hat{a}_{+})[\hat{b}e^{-i[\omega_{M}-(2n-1)\omega_{0}]t}-\hat{b}^{\dagger}e^{i[\omega_{M}-(2n-1)\omega_{0}]t}].
\end{eqnarray}
We choose a  proper $\omega_{0}$ such that one of the underlined terms (denoting as the $n_{0}$ term with the oscillating frequency $\delta=\omega_{M}-2n_{0}\omega_{0}$) is resonant or near-resonant. Then, under the condition~(\ref{SMRWAcond}), the Hamiltonian $\hat{\tilde{H}}(t)$ can be approximated by
\begin{eqnarray}
\hat{H}_{\textrm{approx}}(t)=\frac{g_{0}}{2}J_{2n_{0}}(2\xi)(\hat{a}_{+}^{\dagger}\hat{a}_{-}+\hat{a}_{-}^{\dagger}\hat{a}_{+})
(\hat{b}e^{-i\delta t}+\hat{b}^{\dagger}e^{i\delta t}),
\end{eqnarray}
which is right the approximate Hamiltonian $\hat{H}_{\text{RWA}}(t)$ in Eq.~(\ref{Heff}) under the the relation in Eq.~(\ref{relationsymbare}).

\subsection{B. The physical picture of mechanical displacement enhancement}

\begin{figure}[tbp]
\center
\includegraphics[bb=25 7 393 245, width=1 \textwidth]{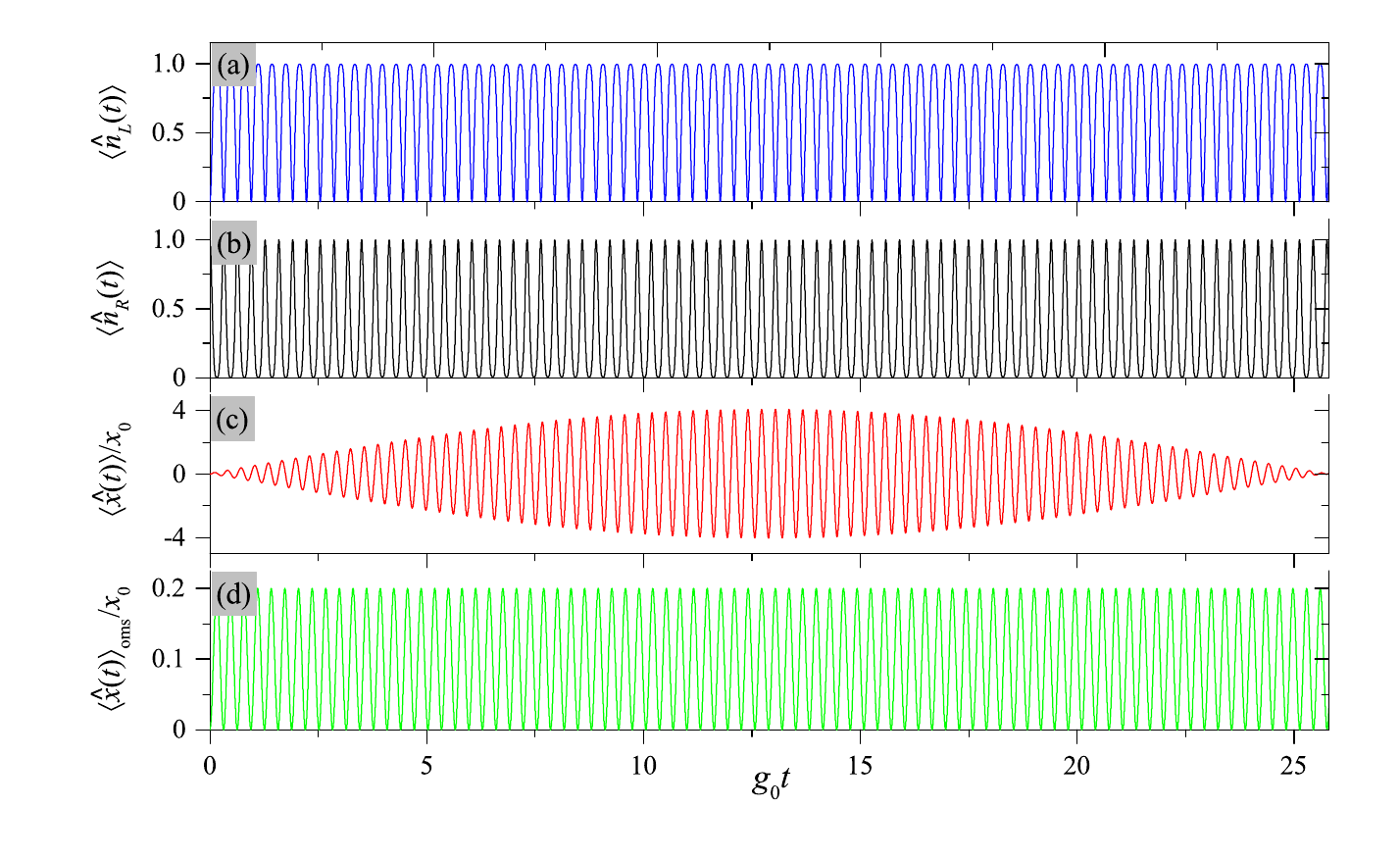}
\caption{(Color online) Time dependence of (a) the average photon number $\langle\hat{n}_{L}(t)\rangle$ in the left cavity, (b) the average photon number $\langle\hat{n}_{R}(t)\rangle$ in the right cavity, and (c) the dimensionless mechanical displacement $\langle\hat{x}(t)\rangle/x_{0}$ in Eq.~(\ref{dynamcsopt}) for the case of modulated two-mode optomecanics. (d) The dynamics of the dimensionless mechanical displacement $\langle\hat{x}(t)\rangle_{\textrm{oms}}/x_{0}$ in Eq.~(\ref{dynamicsxnb}) for the case of single-mode optomechanics. The parameters used are: $\omega_{M}/g_{0}=20$, $\xi=1.5271$, $n_{0}=1$, $\delta=g=g_{0}J_{2n_{0}}(2\xi)/2$, and $\omega_{0}=(\omega_{M}-\delta)/(2n_{0})$. The initial states of the modulated two-mode optomechanical system and the single-mode optomechanical system are $|0\rangle_{L}|1\rangle_{R}|0\rangle_{M}$ and $|1\rangle_{R}|0\rangle_{M}$, respectively. Here we use $g_{0}t$ as the scaled time for keeping the consistency in the two systems.}
\label{FigS1}
\end{figure}
In this section, we study the physical picture of mechanical displacement enhancement induced by the modulated photon hopping in the two-mode optomechanical system, which is described by Hamiltonian~(\ref{fullHamilt}). We will focus on the dynamics of the mechanical displacement created by a single photon. To this end, we write a general single-photon state of the system as
\begin{equation}
\vert\Psi(t)\rangle=\sum_{m=0}^{\infty}[A_{m}(t)\vert 1\rangle_{L}\vert 0\rangle_{R}+B_{m}(t)\vert 0\rangle_{L}\vert 1\rangle _{R}]\vert m\rangle_{M},\label{exactstaPsi}
\end{equation}
where $A_{m}(t)$ and $B_{m}(t)$ are the probability amplitudes of the bases $\vert 1\rangle_{L}\vert 0\rangle_{R}\vert m\rangle_{M}$ and $\vert 0\rangle_{L}\vert 1\rangle _{R}\vert m\rangle_{M}$, respectively.
Based on the full Hamiltonian $\hat{H}(t)$ in Eq.~(\ref{fullHamilt}), the state $\vert\Psi(t)\rangle$ in Eq.~(\ref{exactstaPsi}), and the Schr\"{o}dinger equation
\begin{equation}
i\frac{\partial }{\partial t}\vert\Psi(t)\rangle=\hat{H}(t)\vert\Psi(t)\rangle,
\end{equation}
we can obtain the equations of motion for the probability amplitudes $A_{m}(t)$ and $B_{m}(t)$ as
\begin{subequations}
\label{eqofmotAmBmclo}
\begin{align}
\dot{A}_{m}(t)&=-i(\omega_{c}+m\omega_{M})A_{m}(t)+i\xi \omega_{0}\cos(\omega_{0}t)B_{m}(t),\\
\dot{B}_{m}(t)&=-i(\omega_{c}+m\omega_{M})B_{m}(t)+i\xi \omega_{0}\cos(\omega_{0}t)A_{m}(t)
+ig_{0}[\sqrt{m+1}B_{m+1}(t)+\sqrt{m}B_{m-1}(t)].
\end{align}
\end{subequations}
For studying the mechanical effect of a single photon, we consider the initial state $|0\rangle_{L}|1\rangle_{R}|0\rangle_{M}$ (for comparison purpose with the case of single-mode optomechanics). In principle, if we know the solutions corresponding to the two initial states $|0\rangle_{L}|1\rangle_{R}|0\rangle_{M}$ and $|1\rangle_{L}|0\rangle_{R}|0\rangle_{M}$, then the solution corresponding to the initial state $(1/\sqrt{2})(\vert 1\rangle_{L}\vert 0\rangle_{R}+\vert0\rangle_{L}\vert 1\rangle_{R})\vert 0\rangle_{M}$ in the next section (for generation of mechanical superposition states) can be obtained by superposition. Using the initial state $|0\rangle_{L}|1\rangle_{R}|0\rangle_{M}$, the approximate Hamiltonian $\hat{H}_{\text{RWA}}(t)$, and the transformation $\hat{T}(t)$, we obtain the analytical state of the system as
\begin{equation}
\vert\phi(t)\rangle=\frac{e^{i\vartheta(t)}}{\sqrt{2}}[i\sin[\xi\sin(\omega_{0}t)]\vert 1\rangle_{L}\vert 0\rangle_{R}+\cos[\xi\sin(\omega_{0}t)]\vert 0\rangle_{L}\vert 1\rangle_{R}]|-\beta(t)\rangle_{M},\label{staapproxopt},
\end{equation}
where
\begin{subequations}
\label{thetaandbeat}
\begin{align}
\vartheta(t)&=-\left(\omega_{c}-\frac{g^{2}}{\delta}\right)t-\left(\frac{g}{\delta}\right)^{2}\sin(\delta t),\\
\beta(t)&=-\frac{2ig}{\delta}\sin(\delta t/2)e^{-i(\omega_{M}-\delta/2)t}.
\end{align}
\end{subequations}
We can also obtain the exact state $|\Psi(t)\rangle$ of the system by numerically solving the equations of motion (\ref{eqofmotAmBmclo}) under the initial condition\begin{equation}
B_{0}(0)=1,\hspace{0.5 cm}A_{m\geq0}(0)=B_{m>0}(0)=0.\label{inicondsinleftcav}
\end{equation}
Based on the numerical result, we calculate the dynamics of the observables such as the average photon numbers in the two cavities $\langle\hat{n}_{L}(t)\rangle$ and $\langle\hat{n}_{R}(t)\rangle$, and the dimensionless mechanical displacement $\langle\hat{x}(t)\rangle/x_{0}$ ($x_{0}$ is the zero-point fluctuation of the mechanical resonator):
\begin{subequations}
\label{dynamcsopt}
\begin{align}
\langle\hat{n}_{L}(t)\rangle&=\langle\hat{a}_{L}^{\dagger}\hat{a}_{L}(t)\rangle=\sum_{m=0}^{\infty}\vert A_{m}(t)\vert ^{2},\\
\langle\hat{n}_{R}(t)\rangle&=\langle\hat{a}_{R}^{\dagger}\hat{a}_{R}(t)\rangle=\sum_{m=0}^{\infty}\vert B_{m}(t)\vert ^{2},\\
\langle\hat{x}(t)\rangle/x_{0}&=\langle(\hat{b}^{\dagger}+\hat{b})(t)\rangle=\sum_{n=0}^{\infty }\sqrt{n+1}
[A_{n}^{\ast}(t)A_{n+1}(t)+B_{n}^{\ast}(t)B_{n+1}(t)+A_{n}(t)A_{n+1}^{\ast}(t)+B_{n}(t)B_{n+1}^{\ast}(t)].
\end{align}
\end{subequations}

In Figs.~\ref{FigS1}(a) and~\ref{FigS1}(b), we plot the dynamics of the average photon numbers $\langle\hat{n}_{L}(t)\rangle$ and $\langle\hat{n}_{R}(t)\rangle$ in the left and right cavities. We can see that the single photon hops periodically between the two cavities.
By neglecting the back action of the mechanical oscillation on the right cavity field (i.e., let $g_{0}=0$), an approximate expression of the average photon number in the right cavity can be obtained as
\begin{equation}
\langle\hat{n}_{R}(t)\rangle\approx\cos^{2}[\xi\sin(\omega_{0}t)],\label{phononinrigtcav}
\end{equation}
which is a periodic function with period $T=\pi/\omega_{0}=2\pi/(\omega_{M}-\delta)$. Notice that the mentioned back action will induce
parametric processes involving the photons and phonons, as shown in Eq.~(\ref{Htransall}).

Equation~(\ref{phononinrigtcav}) can be expanded as a summation of sinusoidal functions with frequencies $m\omega_{0}$ ($m$ being integers), and we can understand that there are many periodic driving forces acting on the mechanical resonator. One of the driving forces is near-resonant to the mechanical oscillation, and the near-resonant driving force will increase the displacement amplitude of the mechanical resonator. Other drivings can be neglected approximately because they are largely detuned from the mechanical oscillation under the condition $\omega_{0}\gg g_{0}$. The modulation sidebands can also be designed to suppress other parametric processes and hence a pure and enhanced radiation-pressure interaction can be obtained. Physically, by hopping the single photon into and out of the right cavity (repeating many times) at proper time, the mechanical effect of the single photon will be amplified because the forced displacement effect can be accumulated when the driving force and the mechanical oscillation are in phase in the period of $0$ - $\pi/\delta$ (at time $t=0$, the single photon in the right cavity, and the mechanical resonator is in its ground state). Hence the oscillation amplitude of the resonator increases in this period ($0$ - $\pi/\delta$), as shown in Fig.~\ref{FigS1}(c). We note that the duration $g_{0}t\approx 0$ - $25.8$ corresponds to $\delta t=0$ - $2\pi$ according to the relation $\delta=g=g_{0}J_{2}(2\times1.5271)/2\approx 0.2432\times g_{0}$. After the time $t=\pi/\delta\approx 12.9/g_{0}$, where $\delta$ is the detuning between the frequency of the near-resonant driving force and the mechanical frequency, the  driving force and the mechanical oscillation becomes out phase (during an evolution of time $\pi/\delta$, a phase difference of $\pi$ is generated between the force and the oscillation), and then the oscillation displacement decreases gradually in the period of $\pi/\delta$ - $2\pi/\delta$. The driving of the near-resonant force can be described by the approximate Hamiltonian $\hat{H}_{\text{RWA}}(t)$, and the oscillation of the mechanical mode can be approximately described by $\langle\hat{x}(t)\rangle/x_{0}=(4g/\delta)\sin(\delta t/2)\sin[(\omega_{M}-\delta/2)t]$, which is a sine function oscillation $\sin[(\omega_{M}-\delta/2)t]$ modulated by a sine function envelop $\sin(\delta t/2)$.

As a comparison, below we consider the single-mode optomechanical system, which is described by the Hamiltonian
\begin{equation}
\hat{H}_{\textrm{oms}}=\omega_{c}\hat{a}_{R}^{\dagger}\hat{a}_{R}
+\omega_{M}\hat{b}^{\dagger}\hat{b}-g_{0}\hat{a}_{R}^{\dagger}\hat{a}_{R}(\hat{b}+\hat{b}^{\dagger}),\label{Hamiltoms}
\end{equation}
where $\hat{a}_{R}$ and $\hat{b}$ are the annihilation operators of the single-mode cavity field and the mechanical mode, with the resonance frequencies $\omega_{c}$ and $\omega_{M}$, respectively. The parameter $g_{0}$ is the single-photon optomechanical-coupling strength between the cavity field and the mechanical mode. We can understand that this model is obtained by turning off the photon hopping in the two-mode optomechanical model.

We assume that the single-mode optomechanical system is initially prepared in state $|1\rangle_{R}|0\rangle_{M}$ (corresponding to the case of the modulated two-mode optomechanics), and then the dynamics of the dimensionless mechanical displacement can be obtained as
\begin{eqnarray}
\langle\hat{x}(t)\rangle_{\textrm{oms}}/x_{0}=\langle(\hat{b}+\hat{b}^{\dagger})(t)\rangle
=\frac{4g_{0}}{\omega_{M}}\sin^{2}\left(\frac{\omega_{M}t}{2}\right).\label{dynamicsxnb}
\end{eqnarray}
In Fig.~\ref{FigS1}(d), we plot the dimensionless mechanical displacement $\langle\hat{x}(t)\rangle_{\textrm{oms}}/x_{0}$ as a function of time. We can see that the displacement is a periodic function of time with the period $T=2\pi/\omega_{M}$. Since the single photon always stays in the cavity,  the equilibrium position of the mechanical oscillator is shifted to $x_{0}(2g_{0}/\omega_{M})$. In addition, the oscillation amplitude of the mechanical resonator is $x_{0}(2g_{0}/\omega_{M})$, which is much smaller than the zero-point fluctuation $x_{0}$ because of $g_{0}\ll \omega_{M}$. By comparing the amplitudes in Figs.~\ref{FigS1}(c) and~\ref{FigS1}(d), we can see that the mechanical displacement is largely enhanced by the photon-hopping modulation.

\section{II. Detailed calculations in the closed-system case}

In this section, we present the detailed calculations in the closed-system case. We check the condition~(\ref{SMRWAcond}) under which the RWA performed in obtaining $\hat{H}_{\textrm{RWA}}(t)$ is justified. Concretely, we calculate the fidelity between the approximate state and the exact state, which are governed by the approximate Hamiltonian $\hat{H}_{\textrm{RWA}}(t)$ and the full Hamiltonian $\hat{H}(t)$, respectively. We consider the closed-system case such that the fidelity is only affected by the quality of the RWA.
In addition, we study the quantum interference and coherence effects in the generated mechanical superposition states.

\subsection{A. The approximate states}
\begin{figure}[tbp]
\center
\includegraphics[bb=5 1 531 257, width=0.6 \textwidth]{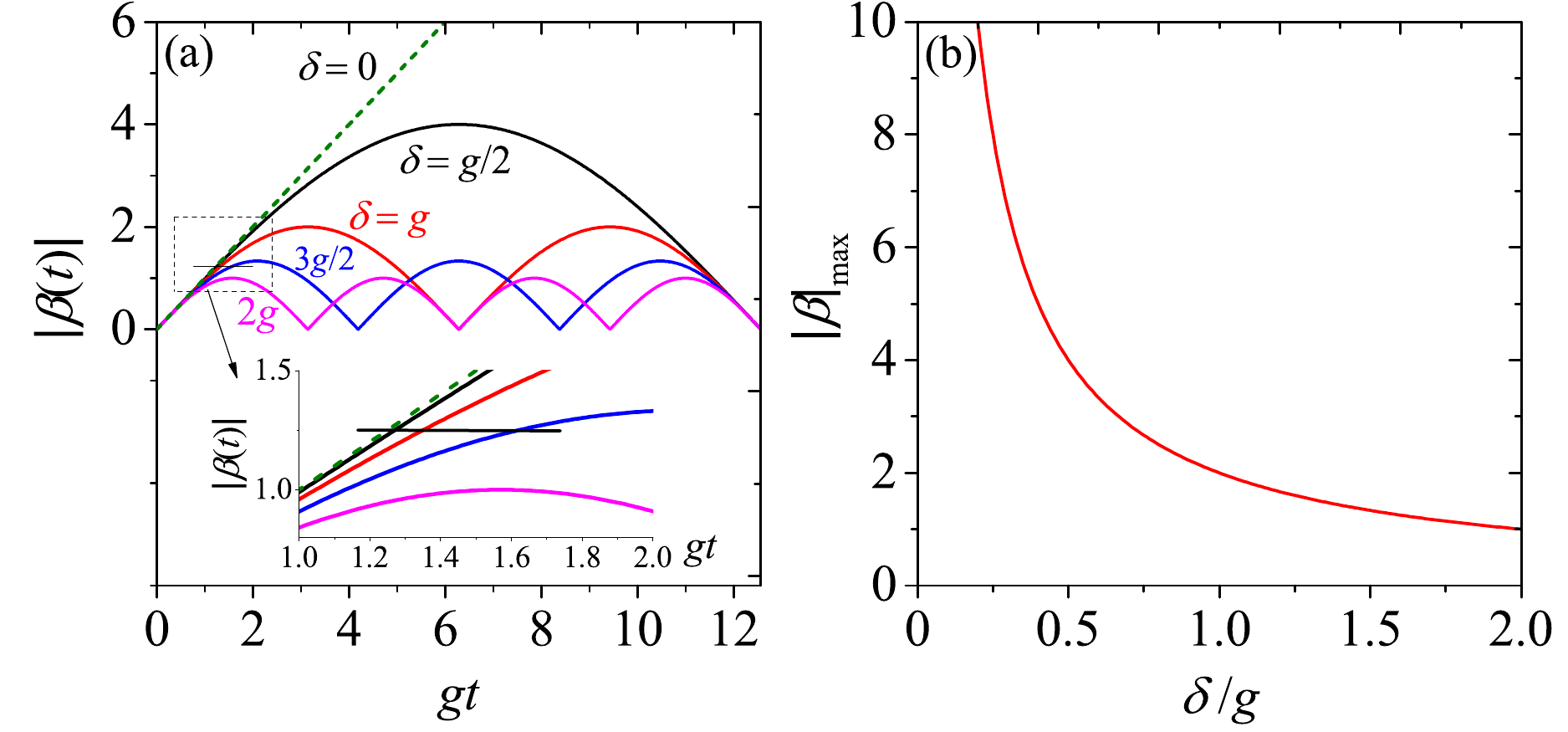}
\caption{(Color online) (a) The displacement amplitude $|\beta(t)|$ versus the scaled time $gt$ at selected values of the detuning $\delta$: $\delta=0$, $g/2$, $g$, $3g/2$, and $2g$. The points of intersection in the inset correspond to the shortest times for obtaining the same displacement ($|\beta(t)|=1.25$ as an example) at different detunings. (b) The maximum displacement $|\beta|_{\textrm{max}}=2g/|\delta|$ as a function of the ratio $\delta/g$.}
\label{FigS2}
\end{figure}
Corresponding to the initial state $\vert\psi(0)\rangle=(1/\sqrt{2})(\vert 1\rangle_{L}\vert 0\rangle_{R}+\vert0\rangle_{L}\vert 1\rangle_{R})\vert 0\rangle_{M}$, the approximate analytical state of the system at time $t$ is
\begin{equation}
\vert\psi(t)\rangle=\frac{e^{i\vartheta(t)}}{\sqrt{2}}[\vert 1\rangle_{L}\vert 0\rangle_{R}\vert\varphi_{L}(t)\rangle_{M}+\vert 0\rangle_{L}\vert 1\rangle_{R}\vert\varphi_{R}(t)\rangle_{M}],\label{staapproxopt}
\end{equation}
where $\vartheta(t)$ is a global phase factor given in Eq.~(\ref{thetaandbeat}).
The two states
\begin{subequations}
\label{varphiLRapp}
\begin{align}
\vert\varphi_{L}(t)\rangle_{M}&=\cos(\mu/2)\vert \beta(t)\rangle_{M}+i\sin(\mu/2)\vert -\beta(t)\rangle_{M},\\
\vert\varphi_{R}(t)\rangle_{M}&=\cos(\mu/2)\vert -\beta(t)\rangle_{M}+i\sin(\mu/2)\vert \beta(t)\rangle_{M},
\end{align}
\end{subequations}
are in quantum superposition of coherent states $\vert \pm\beta(t)\rangle_{M}$, where $\beta(t)$ is given in Eq.~(\ref{thetaandbeat}) and
\begin{equation}
\mu(t)=2\xi\sin(\omega_{0}t).
\end{equation}
In the resonant case $\delta=0$, the displacement becomes
\begin{equation}
\beta_{\textrm{res}}(t)=-igte^{-i\omega_{M}t},
\end{equation}
which grows linearly with time until it becomes so large that the Hamiltonian~(\ref{fullHamilt}) breaks down.

The mechanical displacement amplitude $|\beta(t)|$ and the superposition probability amplitudes $\cos(\mu/2)$ and $\sin(\mu/2)$ are determined by the detuning $\delta$ and the detection time. In Fig.~\ref{FigS2}(a), we plot the displacement amplitude $|\beta(t)|=2g|\sin(\delta t/2)|/|\delta|$ as a function of the scaled time $gt$ at selected values of the detuning $\delta$ (hereafter we assume $\delta\geq0$ for simplicity). We can see that, for a smaller $\delta$, a larger peak displacement can be obtained at proper times (the time corresponding to the first peak value is $t_{0}=\pi/\delta$). In the resonant case $\delta=0$, the amplitude is a linear function of $t$. It should be pointed out that, for obtaining a given displacement, the time $t_{0}=\pi/\delta$ is not the shortest target time. This can be seen in the inset of panel (a). Here we show the shortest times (the intersection points) for the mechanical mode to reach the same displacement ($|\beta(t)|=1.25$ as an example) under different detunings. We can see the resonant driving case is the fastest way to reach the displacement. For example, if we want to obtain a displacement of $2g/\delta$, the time for the resonant case is $t_{\textrm{res}}=2/\delta$, which is shorter than $t_{0}=\pi/\delta$ for the case of $\delta=g$. In our simulations, we consider a general detuning case for numerical convenience. Comparing to the resonant case, the displacement in a general detuning case is less sensitive to the change of the evolution time around $t_{0}$, and hence the truncation is more stable. In Fig.~\ref{FigS2}(b), we show the dependence of the maximum displacement $|\beta|_{\textrm{max}}=2g/|\delta|$ as a function of the ratio $\delta/g$. It shows that the maximum displacement increases with the decreasing of the value of $\delta/g$. For generation of mechanical superposition states with macroscopically distinguishable superposition components, the amplitude $|\beta|$ should be larger than $1$ such that $|\langle\beta|-\beta\rangle| \ll 1$.

\subsection{B. The exact states}
\begin{figure}[tbp]
\center
\includegraphics[bb=1 1 511 281, width=0.9 \textwidth]{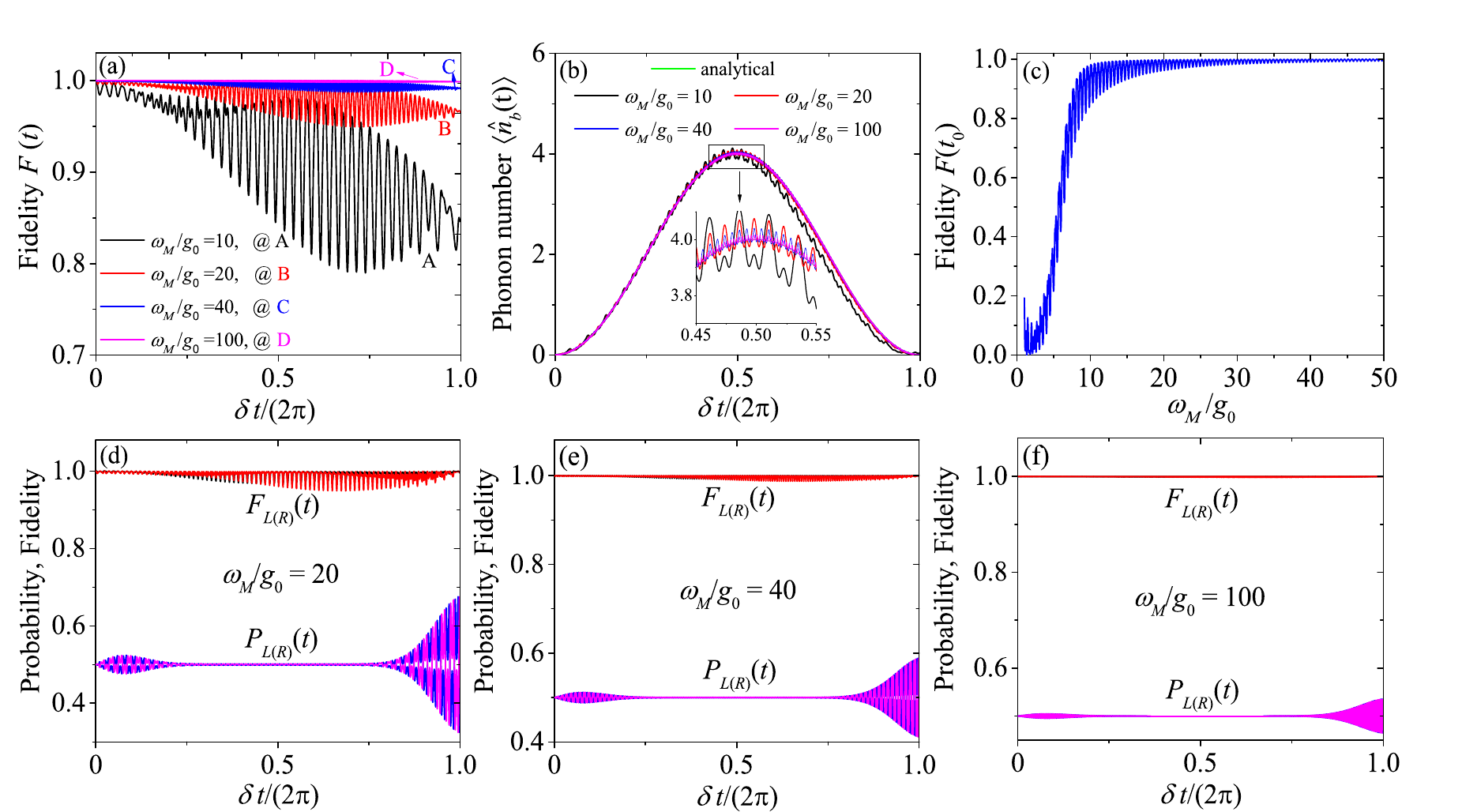}
\caption{(Color online) (a) The fidelity $F(t)$ and (b) the average phonon number $\langle \hat{n}_{b}(t)\rangle$ versus the time $\delta t$ at selected values of the ratio $\omega_{M}/g_{0}$. The analytical result $|\beta(t)|^{2}$ under the RWA is also plotted in (b). (c) The fidelity $F(t_{0})$ at time $t_{0}=\pi/\delta$ versus the ratio $\omega_{M}/g_{0}$. (d-f) The probability $P_{L(R)}(t)$ and fidelity $F_{L(R)}(t)$ of the mechanical states as functions of $\delta t$ when $\omega_{M}/g_{0}=20$, $40$, and $100$, respectively. Other parameters are $n_{0}=1$, $\xi=1.5271$, and $\delta=g$.}
\label{FigS3}
\end{figure}

The exact state of the system can be calculated by solving the equations of motion (\ref{eqofmotAmBmclo}) under a given initial condition. For the initial state $\vert\psi(0)\rangle=\frac{1}{\sqrt{2}}(\vert 1\rangle_{L}\vert 0\rangle_{R}+\vert0\rangle_{L}\vert 1\rangle_{R})\vert 0\rangle_{M}$, the corresponding initial condition is
\begin{equation}
A_{0}(0)=B_{0}(0)=1/\sqrt{2}, \hspace{0.5 cm}A_{m>0}(0)=B_{m>0}(0)=0.
\end{equation}
Under this initial condition, we can numerically solve the equations of motion (\ref{eqofmotAmBmclo}) and obtain the exact state $\vert\Psi(t)\rangle$.

\subsection{C. The fidelities and the validity of the rotating-wave approximation}

The validity of the RWA performed for obtaining Eq.~(\ref{Heff}) can be evaluated by calculating the fidelity $F(t)=\vert\langle\Psi(t)\vert\psi(t)\rangle\vert^{2}$ between the approximate state $\vert\psi(t)\rangle$ and the exact state $\vert\Psi(t)\rangle$. In terms of Eqs.~(\ref{staapproxopt}) and (\ref{exactstaPsi}), the fidelity can be obtained as
\begin{eqnarray}
F(t)&=&\frac{e^{-\vert\beta(t)\vert^{2}}}{2}
\left\vert \sum_{m=0}^{\infty }\left(
\{A_{m}^{\ast}(t) \cos[\xi\sin(\omega_{0}t)]+B_{m}^{\ast}(t)i\sin[\xi\sin
(\omega_{0}t)]\}\frac{[\beta(t)]^{m}}{\sqrt{m!}}\right.\right.\nonumber\\
&&\left.\left.+\{B_{m}^{\ast}(t)\cos[\xi\sin(\omega_{0}t)]+A_{m}^{\ast}(t)i\sin[\xi\sin(\omega_{0}t)]\}\frac{[ -\beta(
t)]^{m}}{\sqrt{m!}}
\right)\right\vert^{2}.
\end{eqnarray}

In Fig.~\ref{FigS3}(a), we plot the fidelity $F(t)$ as a function of the evolution time $t$ when $\omega_{M}/g_{0}$ takes various values and the condition $\delta=g$ is satisfied. We can see that the fidelity $F(t)$ is higher for a larger value of $\omega_{M}/g_{0}$, which is in accordance with the condition~(\ref{SMRWAcond}). The fast oscillation of the fidelity is due to the high-frequency oscillating terms in the full Hamiltonian $\hat{\tilde{H}}(t)$. In Fig.~\ref{FigS3}(b), we show the time dependence of the phonon number $\langle \hat{n}_{b}(t)\rangle=\langle \hat{b}^{\dag}\hat{b}(t)\rangle=\sum_{m=0}^{\infty}m(|A_{m}(t)|^{2}+|B_{m}(t)|^{2})$ when $\omega_{M}$ takes the corresponding values in Fig.~\ref{FigS3}(a). These results are compared with the analytical result $|\beta(t)|^{2}$ (the curve almost overlaps with the curve of $\omega_{M}/g_{0}=100$). It can
be seen that the exact numerical results match better with the approximate analytical result for a larger value of $\omega_{M}/g_{0}$. At time $t_{0}=\pi/\delta$, the phonon number $\langle \hat{n}_{b}\rangle$ reaches its first peak value. To see how the fidelity $F(t_{0})$ at time $t_{0}$ depends on the mechanical frequency, in Fig.~\ref{FigS3}(c) we plot the fidelity $F(t_{0})$ as a function of the ratio $\omega_{M}/g_{0}$ under the relation $\delta=g$. Our result shows that the fidelity $F(t_{0})$ exhibits fast oscillation and its envelope increases quickly with $\omega_{M}$, which agrees with the condition~(\ref{SMRWAcond}).
\begin{figure}[tbp]
\center
\includegraphics[bb=13 2 510 155, width=0.9 \textwidth]{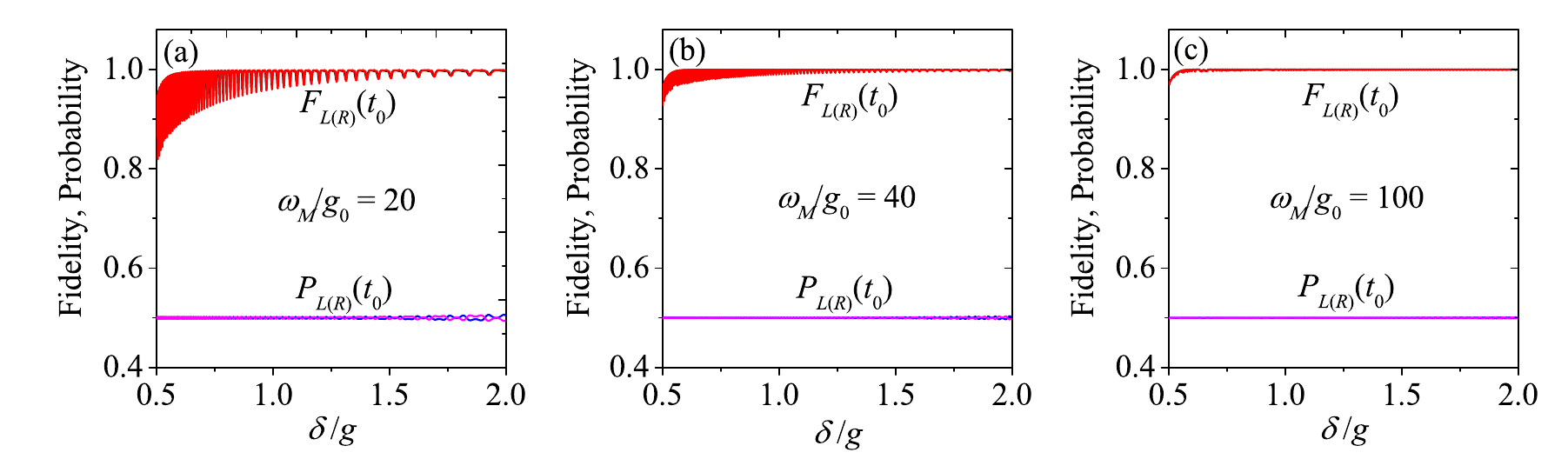}
\caption{(Color online) The probability $P_{L(R)}(t_{0})$ and the fidelity $F_{L(R)}(t_{0})$ at time $t_{0}=\pi/\delta$ as functions of $\delta/g$ when (a ) $\omega_{M}/g_{0}=20$, (b) $\omega_{M}/g_{0}=40$, and (c) $\omega_{M}/g_{0}=100$. Other parameters are $n_{0}=1$ and $\xi=1.5271$.}
\label{FigS4}
\end{figure}

For the state $\vert\Psi(t)\rangle$, when the single photon is detected in the left and right cavities, the
mechanical mode will collapse, respectively, into the states
\begin{eqnarray}
\vert\Psi_{L}(t)\rangle=P_{L}^{-1/2}(t)
\sum_{m=0}^{\infty }A_{m}(t)\vert m\rangle_{M},\hspace{1 cm}
\vert\Psi_{R}(t)\rangle=P_{R}^{-1/2}(t)
\sum_{m=0}^{\infty }B_{m}(t)\vert m\rangle_{M},\label{PsiLRtexactsta}
\end{eqnarray}
where
\begin{eqnarray}
P_{L}(t)=\sum_{m=0}^{\infty }\vert A_{m}(t)\vert^{2}, \hspace{1 cm}
P_{R}(t)=\sum_{m=0}^{\infty }\vert B_{m}(t)\vert ^{2}
\end{eqnarray}
are the probabilities of the single photon to be detected in the left and right cavities, respectively. We also examine the fidelity $F_{s=L(R)}(t)=|\langle\Psi_{s}(t)|\varphi_{s}(t)\rangle|^{2}$ between the target state $|\varphi_{L(R)}(t)\rangle$ and the generated mechanical state $\vert\Psi_{L(R)}(t)\rangle$. Using Eqs.~(\ref{varphiLRapp}) and~(\ref{PsiLRtexactsta}), the fidelities $F_{L}(t)$ and $F_{R}(t)$ can be obtained as
\begin{subequations}
\begin{align}
F_{L}(t)=&\frac{e^{-\vert \beta(t)\vert ^{2}}}{P_{L}(t)}
\left\vert \sum_{m=0}^{\infty }A_{m}^{\ast }(t) \frac{[
\beta(t)]^{m}}{\sqrt{m!}}\left\{\cos[\xi\sin(\omega_{0}t)]+(-1)^{m}i\sin[\xi\sin(\omega_{0}t)]\right\}\right\vert^{2},\\
F_{R}(t)=&\frac{e^{-\vert\beta(t)\vert ^{2}}}{P_{R}(t)}
\left\vert \sum_{m=0}^{\infty }B_{m}^{\ast }(t)\frac{[
\beta(t)]^{m}}{\sqrt{m!}}\left\{i\sin[\xi \sin(\omega_{0}t)]+(-1)^{m}\cos[\xi\sin(\omega_{0}t)]\right\}\right\vert^{2}.
\end{align}
\end{subequations}
In Fig.~\ref{FigS3}(d-f) we show the time dependence of the probabilities $P_{L(R)}(t)$ and the fidelities $F_{L(R)}(t)$ at $\omega_{M}/g_{0}=20$, $40$, and $100$, respectively [when the condition~(\ref{SMRWAcond}) is well satisfied]. In these cases, the probabilities match well with the approximate result of $1/2$ around $t_{0}$ and high fidelities are obtained. For a larger value of the ratio $\omega_{M}/g_{0}$, better fidelities can be obtained.

We also study the dependence of the probability $P_{L(R)}(t_{0})$ and the fidelity $F_{L(R)}(t_{0})$ at time $t_{0}=\pi/\delta$ on the detuning $\delta$. In Fig.~\ref{FigS4}, we display $P_{L(R)}(t_{0})$ and $F_{L(R)}(t_{0})$ as functions of $\delta$ when (a) $\omega_{M}/g_{0}=20$, (b) $\omega_{M}/g_{0}=40$, and (c) $\omega_{M}/g_{0}=100$. Here the detuning $\delta$ is chosen corresponding to $1<|\beta|_{\textrm{max}}<4$. We can see that the probabilities $P_{L}(t_{0})$ and $P_{R}(t_{0})$ are almost independence of the
$\delta$. In contrast, the fidelities $F_{L}(t_{0})$ and $F_{R}(t_{0})$ exhibit minor oscillation with the detuning, and the oscillation amplitude decreases with the increase of $\delta/g$.
The lower envelope of the fidelities becomes worse with the decrease of $\delta/g$. This is because the time $t_{0}=\pi/\delta$ is longer corresponding to a smaller $\delta$, and hence the deviation caused by the approximation increases. For a larger $\omega_{M}/g_{0}$, the fidelities become better.

\subsection{D. The Wigner function and the probability distribution of the rotated quadrature operator}

We can examine the quantum interference and coherence effects in the mechanical states by calculating
the Wigner function and the probability distribution of the rotated quadrature operator.
For a single-mode system described by the density matrix $\hat{\rho}$, the Wigner function is defined by~\cite{Barnettbook}
\begin{eqnarray}
W(\eta)=\frac{2}{\pi}\text{Tr}\left[\hat{D}^{\dagger}(\eta)\hat{\rho}\hat{D}(\eta)(-1)^{\hat{b}^{\dagger}\hat{b}}\right],
\end{eqnarray}
where $\hat{D}(\eta)=\exp(\eta\hat{b}^{\dagger}-\eta^{\ast}\hat{b})$ is the displacement operator.
For the rotated quadrature operator
\begin{eqnarray}
\hat{X}(\theta)=\frac{1}{\sqrt{2}}(\hat{b}e^{-i\theta}+\hat{b}^{\dag}e^{i\theta}),
\end{eqnarray}
we denote the eigenstate as $|X(\theta)\rangle$: $\hat{X}(\theta)|X(\theta)\rangle=X(\theta)|X(\theta)\rangle$, then, for the density matrix $\hat{\rho}(t)$, the probability distribution of the rotated quadrature operator $\hat{X}(\theta)$ is defined by~\cite{Milburnbook}
\begin{equation}
P[X(\theta)]=\langle X(\theta)\vert\hat{\rho}(t)\vert X(\theta)\rangle.
\end{equation}
\begin{figure}[tbp]
\center
\includegraphics[bb=18 0 560 298, width=0.9 \textwidth]{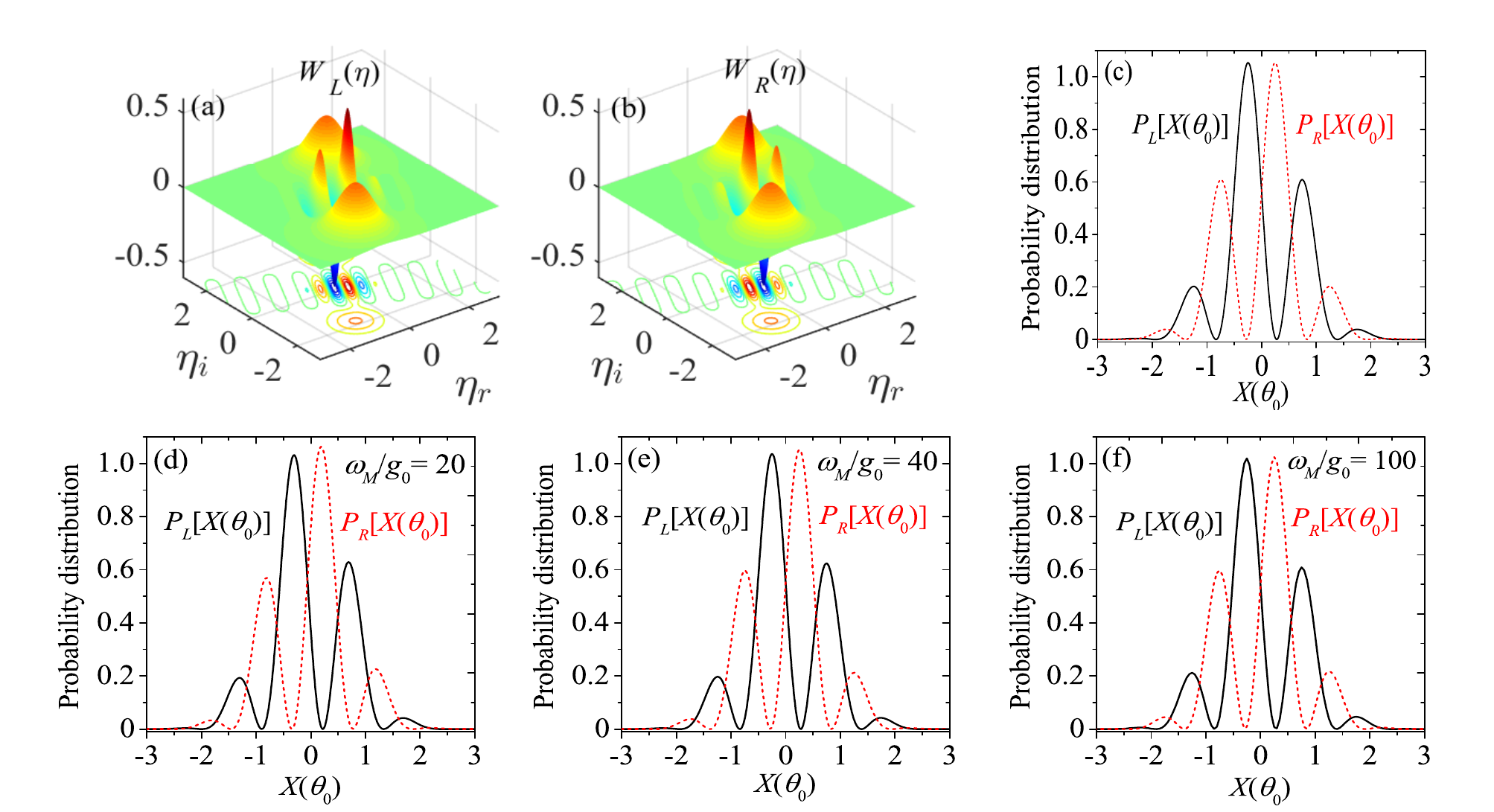}
\caption{(Color online) The Wigner functions (a) $W_{L}(\eta)$ and (b) $W_{R}(\eta)$, and the probability distributions (c) $P_{L(R)}[X(\theta_{0})]$ of the approximate states $\vert\varphi_{L}(t_{d})\rangle_{M}$ and $\vert\varphi_{L}(t_{d})\rangle_{M}$ with $\beta(t_{d})=-0.8878-1.7911i$ ($\omega_{M}/g_{0}=20$). The rotating angle is  $\theta_{0}=\arg[\beta(t_{d})]-\pi/2=-3.6018$. (d-f) The probability distributions $P_{L}[X(\theta_{0})]$ and $P_{R}[X(\theta_{0})]$ of the exact states $\vert\Psi_{L}(t_{d})\rangle_{M}$ and $\vert\Psi_{R}(t_{d})\rangle_{M}$ at selected values of the mechanical frequency: $\omega_{M}/g_{0}=20$, $40$, and $100$. It should be pointed out that corresponding to different values of $\omega_{M}/g_{0}$, the detection time $t_{d}$ and the rotating angle $\theta_{0}$ are different. Other parameters are: $n_{0}=1$, $\xi=1.5271$, and $\delta=g$.}
\label{FigS5}
\end{figure}

For the analytical states $\vert\varphi_{L}(t)\rangle_{M}$ and $\vert\varphi_{R}(t)\rangle_{M}$ given in Eq.~(\ref{varphiLRapp}), the Wigner functions can be obtained as
\begin{subequations}
\begin{align}
W_{L}(\eta)=&\frac{2}{\pi}\cos^{2}[\xi\sin(\omega_{0}t)]\exp[-2\vert\eta-\beta(t)\vert^{2}]+\frac{2}{\pi}\sin^{2}[\xi\sin(\omega_{0}t)]
\exp[-2\vert\eta+\beta(t)\vert^{2}]\notag\\
&-\frac{2}{\pi}\exp(-2\vert\eta\vert^{2})\sin[2\xi\sin(\omega_{0}t)]\sin[4\text{Im}[\eta\beta^{\ast}(t)]],\\
W_{R}(\eta)=&\frac{2}{\pi}\cos^{2}[\xi\sin(\omega_{0}t)]\exp[-2\vert\eta+\beta(t)\vert^{2}]+\frac{2}{\pi}\sin^{2}[\xi\sin(\omega_{0}t)]
\exp[-2\vert\eta-\beta(t)\vert^{2}]\notag\\
&+\frac{2}{\pi}\exp(-2\vert\eta\vert^{2})\sin[2\xi\sin(\omega_{0}t)]\sin[4\text{Im}[\eta\beta^{\ast}(t)]].
\end{align}
\end{subequations}
The probability distributions of the rotated quadrature operator $\hat{X}(\theta)$ for the states $\vert\varphi_{L}(t)\rangle_{M}$ and $\vert\varphi_{R}(t)\rangle_{M}$ are
\begin{subequations}
\begin{align}
P_{L}[X(\theta)]=&\vert\,_{M}\!\langle X(\theta)\vert\varphi_{L}(t)\rangle_{M}\vert^{2}\notag\\
=&e^{-\vert\beta(t)\vert^{2}}\left\vert\sum_{p=0}^{\infty}\{\cos[\xi\sin(\omega_{0}t)]+i(-1)^{p}\sin[\xi\sin(\omega_{0}t)]\}\frac{\beta^{p}(t)
}{\sqrt{p!}}\frac{1}{\sqrt{\pi^{1/2}2^{p}p!}}H_{p}[X(\theta)]e^{-X^{2}(\theta)/2}
e^{-i\theta p}\right\vert^{2},\\
P_{R}[X(\theta)]=&\vert\,_{M}\!\langle X(\theta)\vert\varphi_{R}(t)\rangle_{M}\vert^{2}\notag\\
=&e^{-\vert\beta(t)\vert ^{2}}\left\vert\sum_{p=0}^{\infty}\{i\sin[\xi\sin(\omega_{0}t)]+(-1)^{p}\cos[\xi\sin(\omega_{0}t)]\}\frac{\beta^{p}(t)
}{\sqrt{p!}}\frac{1}{\sqrt{\pi^{1/2}2^{p}p!}}H_{p}[X(\theta)]e^{-X^{2}(\theta)/2}
e^{-i\theta p}\right\vert^{2},
\end{align}
\end{subequations}
where $H_{m}[z]$ are the Hermite polynomials. In the derivation of the probability distributions, we have used the relation
\begin{eqnarray}
\langle X(\theta)\vert n\rangle=\frac{1}{\sqrt{\pi^{1/2}2^{n}n!}}H_{n}[X(\theta)]\exp\left[-\frac{1}{2}X^{2}(\theta)\right]e^{-i\theta n}.
\end{eqnarray}

We now study the properties of the approximate states. As an example, we choose the parameters: $\omega_{M}/g_{0}=20$, $n_{0}=1$, $\xi=1.5271$, $\delta=g$, and $t_{d}=12.6664/g_{0}$. Then the corresponding approximate states become
\begin{subequations}
\begin{align}
\vert\varphi_{L}(t_{d})\rangle_{M}&=(1/\sqrt{2})(\vert \beta(t_{d})\rangle_{M}-i\vert-\beta(t_{d})\rangle_{M}),\\ \vert\varphi_{R}(t_{d})\rangle_{M}&=(1/\sqrt{2})(\vert -\beta(t_{d})\rangle_{M}-i\vert\beta(t_{d})\rangle_{M}),
\end{align}
\end{subequations}
where $\beta(t_{d})=-0.8878-1.7911i$ with argument $\arg[\beta(t_{d})]=-2.0310$ and amplitude $|\beta(t_{d})|\approx 2$.
In Fig.~\ref{FigS5}, we plot the Wigner functions (a) $W_{L}(\eta)$ and (b) $W_{R}(\eta)$, and (c) the probability distributions $P_{L}[X(\theta_{0})]$ and $P_{R}[X(\theta_{0})]$ of the rotated quadrature operator $\hat{X}(\theta_{0})$. We see obvious interference evidence in the Wigner functions, and the positions of the two main peaks are located at $\pm\beta(t_{d})$ in the phase space. The Wigner function $W_{R}(\eta)$ is a rotation of $W_{L}(\eta)$ by $\pi$ about the origin in phase space, which agrees with the analytical relation $\vert\varphi_{R}(t_{d})\rangle_{M}=e^{i\pi \hat{b}^{\dag}\hat{b}}\vert\varphi_{L}(t_{d})\rangle_{M}$. In plotting the probability distributions, we choose the rotation angle as $\theta_{0}=\arg[\beta(t_{d})]-\pi/2=-3.6018$, which means that the quadrature direction is perpendicular to the link line between the locations of the two superposition coherent amplitudes. The interference is maximum in this direction because the two coherent states are projected onto the quadrature such that they overlap exactly. The oscillation in the curves is a distinct evidence of the quantum interference between the superposition state components. At the same time, $P_{L}[X(\theta_{0})]$ and $P_{R}[X(\theta_{0})]$ are symmetry to each other about the vertical axis $X(\theta_{0})=0$, which is also a consequence of the relation $\vert\varphi_{R}(t_{d})\rangle_{M}=e^{i\pi \hat{b}^{\dag}\hat{b}}\vert\varphi_{L}(t_{d})\rangle_{M}$.

It should be pointed out that, in contrast to the approximate states, the probability distributions $P_{L}[X(\theta_{0})]$ and $P_{R}[X(\theta_{0})]$ of the exact states $\vert\Psi_{L}(t)\rangle$ and $\vert\Psi_{R}(t)\rangle$ are not exact symmetry to each other about the vertical axis $X(\theta_{0})=0$ when the ratio $\omega_{M}/g_{0}$ is not large enough. In panels (d-f) of Fig.~\ref{FigS5}, we show $P_{L(R)}[X(\theta_{0})]$ of the exact state $|\Psi_{L(R)}(t_{d})\rangle$ when $\omega_{M}/g_{0}=20$, $40$, and $100$. We can see that with the increase of $\omega_{M}/g_{0}$, the probability distributions $P_{L}[X(\theta_{0})]$ and $P_{R}[X(\theta_{0})]$ become more and more symmetric. This means that the RWA becomes better for a larger value of $\omega_{M}/g_{0}$. In plotting panels (d-f), corresponding to different values of $\omega_{M}/g_{0}$, the detection times $t_{d}$ and the rotating angle $\theta_{0}$ are also different. For $\omega_{M}/g_{0}=20$, we have  $t_{d}=12.6664/g_{0}$ and $\theta_{0}=-3.6018$. For $\omega_{M}/g_{0}=40$, we take $t_{d}=12.8285/g_{0}$ and $\theta_{0}=0.5010$, and for $\omega_{M}/g_{0}=100$, we have  $t_{d}=12.9228/g_{0}$ and $\theta_{0}=0.4895$.

\section{III. Detailed calculations in the open-system case}

\subsection{A. The master equation and its solution}

In this section, we present the detailed calculations in the open-system case, in which the evolution of the system is governed by the quantum master equation
\begin{equation}
\dot{\hat{\rho}}=i[\hat{\rho},\hat{H}(t)]+\gamma_{c}\mathcal{D}[\hat{a}_{L}]\hat{\rho}+\gamma_{c}\mathcal{D}[\hat{a}_{R}]\hat{\rho}
+\gamma_{M}(n_{\textrm{th}}+1)\mathcal{D}[\hat{b}]\hat{\rho}+\gamma_{M}n_{\textrm{th}}\mathcal{D}[\hat{b}^{\dag}]\hat{\rho},\label{mastereq}
\end{equation}
where $\mathcal{D}[\hat{o}]\hat{\rho}=\hat{o}\hat{\rho} \hat{o}^{\dagger}-(\hat{o}^{\dagger}\hat{o}\hat{\rho}+\hat{\rho} \hat{o}^{\dagger}\hat{o})/2$ is the standard Lindblad superoperator for photon and phonon dampings, $\gamma_{c}$ and $\gamma_{M}$ are the damping rates of the cavity fields and the mechanical mode, respectively, and $n_{\textrm{th}}$ is the thermal phonon occupation number. The full Hamiltonian $\hat{H}(t)$ is given by Eq.~(\ref{fullHamilt}).
For convenience, we write the density matrix of the total system in the Fock state space as
\begin{equation}
\hat{\rho}(t)=\sum_{m,j,p,n,k,q=0}^{\infty}\rho_{m,j,p,n,k,q}(t)
\vert m\rangle\!_{L\;L}\!\langle n\vert\otimes\vert j\rangle\!_{R\;R}\!\langle k\vert\otimes\vert p\rangle\!_{M\;M}\!\langle q\vert,
\end{equation}
where $\vert m (n)\rangle_{L}$, $\vert j (k)\rangle_{R}$, and $\vert p (q)\rangle_{M}$ are the Fock states of the left cavity, the right cavity, and the mechanical mode, respectively. By substituting the density matrix $\hat{\rho}(t)$ into the master equation~(\ref{mastereq}), we can obtain the equations of motion of the density matrix elements $\rho_{m,j,p,n,k,q}(t)$ as
\begin{eqnarray}
&&\dot{\rho}_{m,j,p,n,k,q}(t)\nonumber\\
&=&\left\{i\left[(n-m+k-j)\omega_{c}+(q-p)\omega_{M}\right]
-\left[\frac{\gamma_{c}}{2}(m+n+j+k)+\frac{\gamma_{M}}{2}
\left[(2n_{\textrm{th}}+1)(p+q)+2n_{\textrm{th}}\right]\right]
\right\} \rho _{m,j,p,n,k,q}(t)\notag\\
&&-i\xi \omega _{0}\cos(\omega_{0}t)\left[\sqrt{(n+1)k}\rho_{m,j,p,n+1,k-1,q}(t)+\sqrt{n(
k+1)}\rho_{m,j,p,n-1,k+1,q}(t)\right]\notag \\
&&+i\xi \omega _{0}\cos(\omega_{0}t)\left[\sqrt{m(j+1)}\rho_{m-1,j+1,p,n,k,q}(t)+\sqrt{(m+1)j
}\rho_{m+1,j-1,p,n,k,q}(t)\right]\notag \\
&&-ikg_{0}\left[\sqrt{q}\rho_{m,j,p,n,k,q-1}(t)+\sqrt{q+1}
\rho _{m,j,p,n,k,q+1}(t)\right]+ijg_{0}\left[\sqrt{p+1}\rho_{m,j,p+1,n,k,q}(t)+\sqrt{p}
\rho _{m,j,p-1,n,k,q}(t) \right]\notag \\
&&+\gamma _{c}\left[\sqrt{(m+1)(n+1)}\rho_{m+1,j,p,n+1,k,q}(t)+\sqrt{(j+1)(k+1)
}\rho _{m,j+1,p,n,k+1,q}(t)\right]\notag \\
&&+\gamma _{M}\left[\sqrt{(p+1)(q+1)}(n_{\textrm{th}}+1)\rho_{m,j,p+1,n,k,q+1}(t)+\sqrt{qp}n_{\textrm{th}}\rho _{m,j,p-1,n,k,q-1}(t)\right].
\end{eqnarray}
\begin{figure}[tbp]
\center
\includegraphics[bb=20 0 345 273, width=0.35 \textwidth]{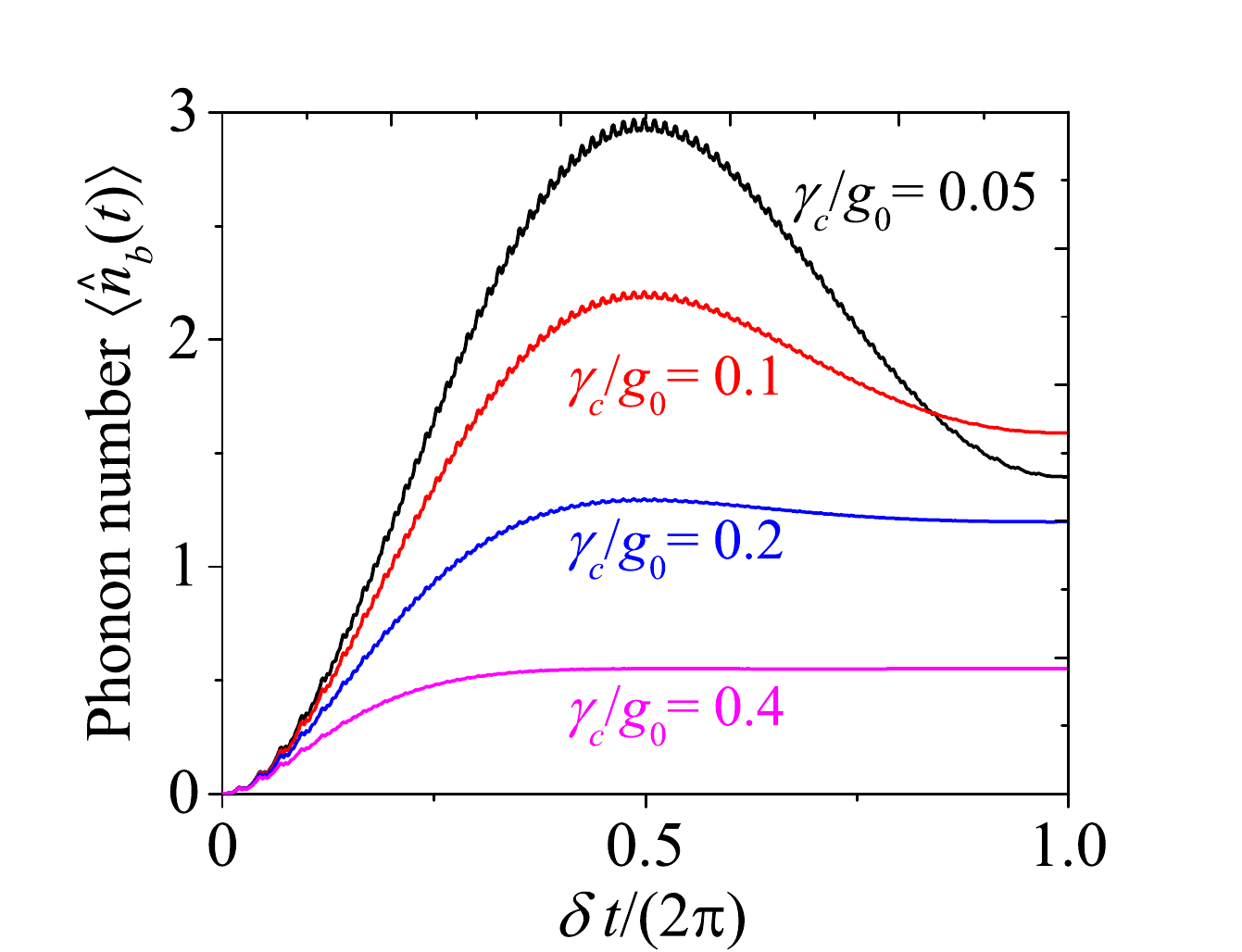}
\caption{(Color online) The time dependence of the average phonon number $\langle\hat{n}_{b}(t)\rangle$ at selected values of the cavity-field decay rate $\gamma_{c}$. Other parameters are $\omega_{M}/g_{0}=20$, $n_{0}=1$, $\xi=1.5271$, $\delta=g$, $\gamma_{M}/g_{0}=0.0001$, and $n_{\textrm{th}}=4$.}
\label{FigS6}
\end{figure}

By solving these equations of motion under the initial condition corresponding to the initial state $\vert\psi(0)\rangle=(1/\sqrt{2})(\vert 1\rangle_{L}\vert 0\rangle_{R}+\vert0\rangle_{L}\vert 1\rangle_{R})\vert 0\rangle_{M}$, we obtain the state evolution of the density matrix $\hat{\rho}(t)$, and then we can calculate the dynamics of the system's observables such as the average phonon number
\begin{eqnarray}
\langle\hat{n}_{b}(t)\rangle=\langle \hat{b}^{\dagger}\hat{b}(t)\rangle=\sum_{m,j,p=0}^{\infty}p\rho_{m,j,p,m,j,p}(t).
\end{eqnarray}
In realistic simulations, the cavity phonon number can be restricted into a subspace with zero photon and one photon. This is because we consider the single photon initial state and the cavity field will not be excited due to the vacuum baths. In addition, the dimension of the phonon numbers needs to be truncated up to a definite number, which is determined by generated states.

In Fig.~\ref{FigS6}, we plot the dynamics of the average phonon number $\langle \hat{n}_{b}(t)\rangle$ at selected values of $\gamma_{c}$. The curves show that, with the increase of $\gamma_{c}$, the maximum average phonon number decreases, and the oscillation disappears gradually. This is because a larger $\gamma_{c}$ corresponds to a faster photon leakage. When the single photon leaks out of the cavity, then the mechanical resonator will experience a dissipative evolution. The photon leakage can be seen directly from the probability $P_{L(R)}(t)$ of a single photon in the left (right) cavity, as shown in the main text.

\subsection{B. The fidelity and probability of the mechanical superposition states}

For generation of mechanical superposition states, we need to measure the state of the cavity fields at a proper time such that the mechanical resonator collapses into the reduced density matrices. Specifically, when the single photon is detected
in the left and right cavities, the corresponding reduced density matrices of the mechanical mode are
\begin{subequations}
\begin{align}
\hat{\rho}_{M}^{(L)}(t)&=\frac{1}{P_{L}(t)}\sum_{p,q=0}^{\infty}\rho_{1,0,p,1,0,q}(t)\vert p\rangle\!_{M\,M}\!\langle q\vert,\\
\hat{\rho}_{M}^{(R)}(t)&=\frac{1}{P_{R}(t)}\sum_{p,q=0}^{\infty}\rho_{0,1,p,0,1,q}(t)\vert p\rangle\!_{M\,M}\!\langle q\vert,
\end{align}
\end{subequations}
where
\begin{equation}
P_{L}(t)=\sum_{p=0}^{\infty }\rho_{1,0,p,1,0,p}(t),\hspace{2 cm} P_{R}(t)=\sum_{p=0}^{\infty }\rho_{0,1,p,0,1,p}(t)
\end{equation}
are the corresponding detection probabilities of the single photon in the left and right cavities, respectively.

The fidelity $F_{s=L(R)}(t)=\langle\varphi_{s}(t)\vert\hat{\rho}_{M}^{(s)}(t)\vert\varphi_{s}(t)\rangle$ between the target state $\vert\varphi_{s}(t)\rangle$ and the generated state $\hat{\rho}_{M}^{(s)}(t)$ can be obtained as
\begin{subequations}
\begin{align}
F_{L}(t)=&\frac{e^{-\vert\beta(t)\vert^{2}}}{P_{L}(t)}\sum_{p,q=0}^{\infty}\rho_{1,0,p,1,0,q}(t)\frac{[\beta
^{\ast}(t)]^{p}\beta^{q}(t)}{\sqrt{p!q!}}\nonumber\\
&\times\left\{\cos ^{2}[\xi\sin(\omega_{0}t)]+(-1)^{p+q}\sin^{2}[\xi\sin(\omega_{0}t)]
+i[(-1)^{q}-(-1)^{p}]\sin[\xi\sin(\omega_{0}t)]\cos[\xi\sin(\omega_{0}t)]\right\},\\
F_{R}(t)=&\frac{e^{-\vert\beta(t)\vert^{2}}}{P_{R}(t)}\sum_{p,q=0}^{\infty}\rho _{0,1,p,0,1,q}(t)\frac{[\beta^{\ast}(t)]^{p}\beta^{q}(t)}{\sqrt{p!q!}}\nonumber\\
&\times\left\{\sin^{2}[\xi\sin(\omega_{0}t)]+(-1)^{p+q}\cos^{2}[\xi\sin(\omega_{0}t)]
+i[(-1)^{p}-(-1)^{q}]\sin[\xi\sin(\omega_{0}t)]\cos[\xi\sin(\omega_{0}t)]\right\}.
\end{align}
\end{subequations}

\subsection{C. The Wigner function and the probability distribution of the rotated quadrature operator}

Similar to the closed-system case, to see the quantum interference and coherence effects in the mechanical superposition states, we calculate the Wigner function and the probability distribution of the rotated quadrature operator. For the mechanical states $\rho_{M}^{(L)}(t)$ and $\rho_{M}^{(R)}(t)$, the Wigner functions can be obtained as
\begin{subequations}
\begin{align}
W_{L}(\eta)=&\frac{2}{\pi}\frac{1}{P_{L}(t)}\sum_{l,p,q=0}^{\infty}(-1)^{l}\rho_{1,0,p,1,0,q}(t)
(_{M}\!\langle p\vert \hat{D}(\eta)\vert l\rangle_{M})^{\ast}\,_{M}\!\langle q\vert \hat{D}(\eta)\vert l\rangle_{M},\\
W_{R}(\eta)=&\frac{2}{\pi}\frac{1}{P_{R}(t)}\sum_{l,p,q=0}^{\infty}(-1)^{l}\rho_{0,1,p,0,1,q}(t)
(_{M}\!\langle p\vert \hat{D}(\eta)\vert l\rangle_{M})^{\ast}\,_{M}\!\langle q\vert \hat{D}(\eta)\vert l\rangle_{M}.
\end{align}
\end{subequations}
Here the matrix elements of the displacement operator in the Fock state space can be calculated by the following relations~\cite{Buzek1990}
\begin{equation}
_{M}\!\langle m\vert\hat{D}(\eta)\vert n\rangle\!_{M}=\left\{
\begin{array}{c}
\sqrt{\frac{m!}{n!}}\exp\left(-\frac{\vert\eta\vert^{2}}{2}\right)(-\eta^{\ast})^{n-m}L_{m}^{n-m}(\vert\eta
\vert ^{2}),\hspace{0.5 cm}n>m, \\
\sqrt{\frac{n!}{m!}}\exp\left(-\frac{\vert\eta\vert^{2}}{2}\right)(\eta)^{m-n}L_{n}^{m-n}(\vert\eta\vert^{2}),\hspace{0.5 cm}m>n.
\end{array}
\right.
\end{equation}
with $L_{n}^{m}(x)$ being the associated Laguerre polynomials.
\begin{figure}[tbp]
\center
\includegraphics[bb=32 0 497 293, width=0.8 \textwidth]{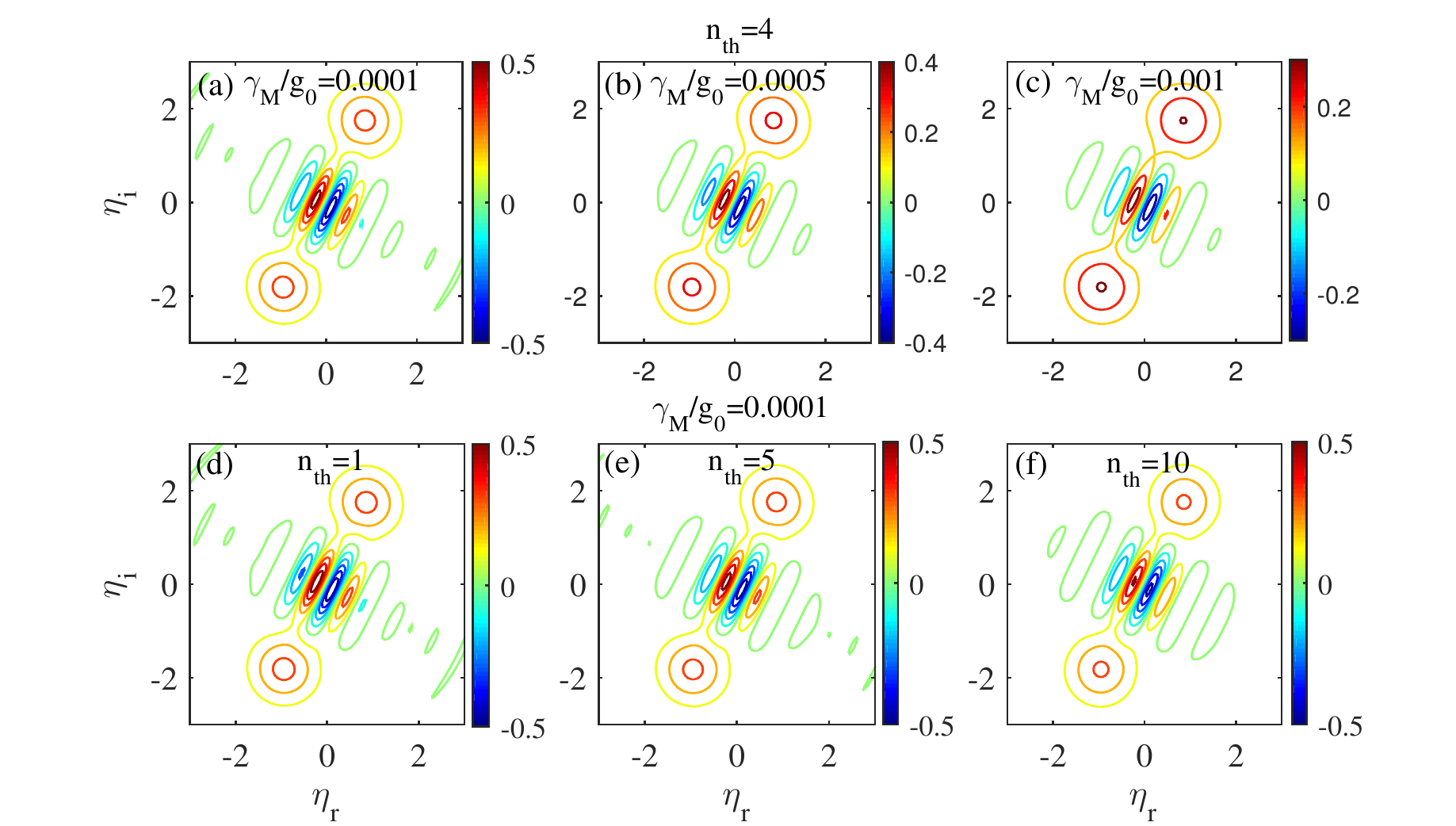}
\caption{(Color online) The Wigner function $W_{L}(\eta)$ of the state $\hat{\rho}_{M}^{(L)}(t_{d})$ at selected values of the mechanical decay rate $\gamma_{M}$ and the thermal phonon occupation number $n_{\textrm{th}}$, corresponding to Fig.~3 in the main text. (a-c) $n_{\textrm{th}}=4$ and $\gamma_{M}/g_{0}=0.0001$, $0.0005$, $0.001$. (d-f) $\gamma_{M}/g_{0}=0.0001$ and $n_{\textrm{th}}=1$, $5$, $10$. Other parameters are $\omega_{M}/g_{0}=20$, $n_{0}=1$, $\xi=1.5271$, $\delta=g$, and $\gamma_{c}/g_{0}=0.2$.}
\label{FigS7}
\end{figure}

For the mechanical states $\rho_{M}^{(L)}(t)$ and $\rho_{M}^{(R)}(t)$, the probability distributions of the rotated quadrature operator $\hat{X}(\theta)$ can be obtained as
\begin{subequations}
\begin{align}
P_{L}[X(\theta)]=&\frac{1}{P_{L}(t)}\sum_{p,q=0}^{\infty }\rho_{1,0,p,1,0,q}(t)\frac{1}{\sqrt{\pi
2^{(p+q)}p!q!}}H_{p}[X(\theta)]H_{q}[X(\theta)]\exp[-X^{2}(\theta)]e^{i\theta(q-p)},\\
P_{R}[X(\theta)]=&\frac{1}{P_{R}(t)}\sum_{p,q=0}^{\infty }\rho_{0,1,p,0,1,q}(t)\frac{1}{\sqrt{\pi2^{(p+q)}p!q!}}H_{p}[X(\theta)]H_{q}[
X(\theta)]\exp[-X^{2}(\theta)]e^{i\theta(q-p)},
\end{align}
\end{subequations}
where $H_{m}[z]$ are the Hermite polynomials.

In Fig.~2 and Fig.~3 of the main text, we studied how the cavity field decay and the mechanical dissipation affect the probability, the fidelity, and the probability distribution of the rotated quadrature operator of the generated mechanical states. We also showed the Wigner function of the mechanical states in the presence of the cavity field decay [Fig.~2(c)]. As a complementary, in Fig.~\ref{FigS7} we show the Wigner functions of the generated mechanical states at selected values of the mechanical dissipation parameters (corresponding to Fig.~3 in the main text). The results show that the locations of the two peaks in the Wigner function do not change with the mechanical dissipations in our simulations. However, the interference pattern attenuates gradually with the increase of $\gamma_{M}$ and $n_{th}$.

\subsection{D. The results in the case of $\omega_{M}/g_{0}=100$}
\begin{figure}[tbp]
\center
\includegraphics[bb=0 0 575 225, width=1 \textwidth]{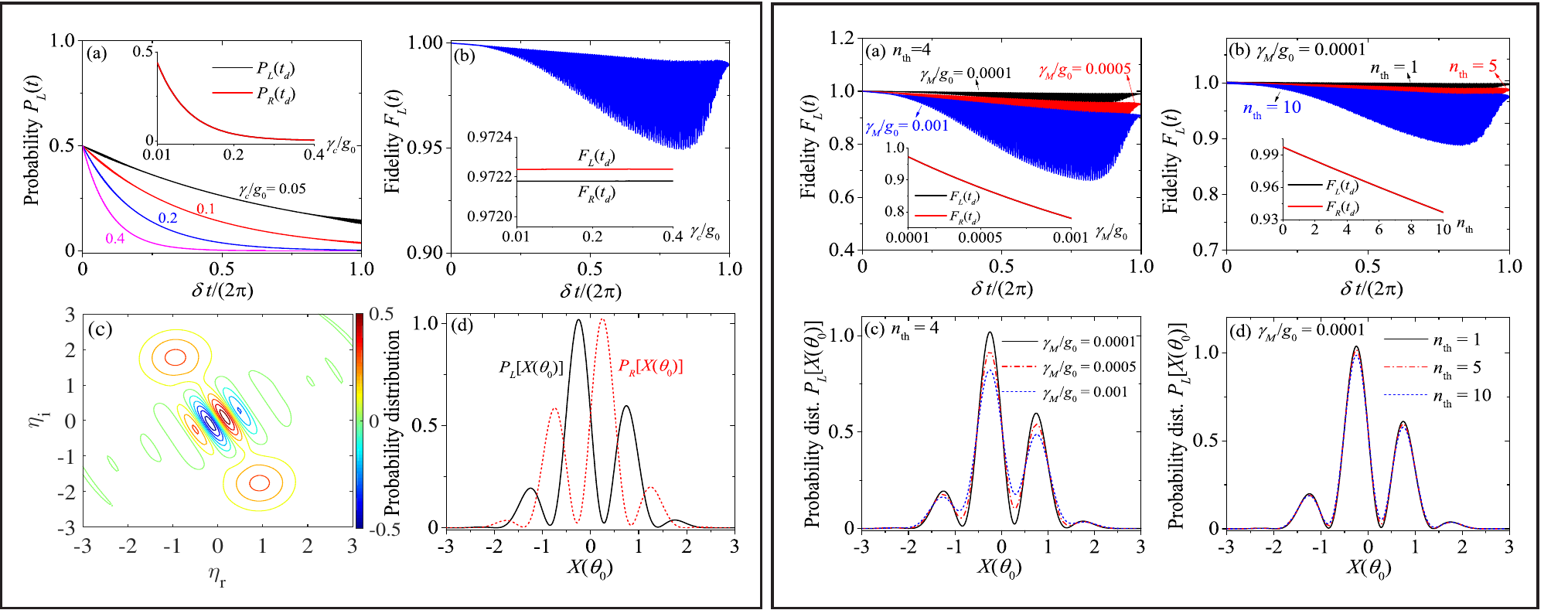}
\caption{(Color online) Figures in the left and right boxes correspond respectively to Fig.~2 and Fig.~3 in the main text. The only difference is that the mechanical frequency used in these figures is $\omega_{M}/g_{0}=100$, while the ratio $\omega_{M}/g_{0}=20$ is used in Fig.~2 and Fig.~3 in the main text. \emph{Left box}: (a) The probability $P_{L}(t)$ and (b) the fidelity $F_{L}(t)$ versus $\delta t$ at selected values of the cavity-field decay rate $\gamma_{c}$ [insets: the probability $P_{L(R)}(t_{d})$ and the fidelity $F_{L(R)}(t_{d})$ at time $t_{d}$ versus $\gamma_{c}/g_{0}$]. (c) The Wigner function $W_{L}(\eta)$ (with $\eta=\eta_{r}+i\eta_{i}$) and (d) the probability distribution of the rotated quadrature operator $P_{L}[X(\theta_{0})]$ for the state $\hat{\rho}^{(L)}_{M}(t_{d})$. Other parameters are $\omega_{M}/g_{0}=100$, $n_{0}=1$, $\xi=1.5271$, $\delta=g$, $\gamma_{M}/g_{0}=0.0001$, and $n_{\textrm{th}}=4$. \emph{Right box}: The fidelity $F_{L}(t)$ versus $\delta t$ at selected values of (a) the mechanical decay rate $\gamma_{M}/g_{0}$ and (b) the thermal phonon occupation number $n_{\textrm{th}}$ [insets: the fidelity $F_{L(R)}(t_{d})$ at time $t_{d}$ versus $\gamma_{M}/g_{0}$ and $n_{\textrm{th}}$]. The probability distribution of the rotated quadrature operator $P_{L}[X(\theta_{0})]$ of the state $\hat{\rho}^{(L)}_{M}(t_{d})$ at selected values of (c) $\gamma_{M}/g_{0}$ and (d) $n_{\textrm{th}}$. Other parameters are $\omega_{M}/g_{0}=100$, $n_{0}=1$, $\xi=1.5271$, $\delta=g$, and $\gamma_{c}/g_{0}=0.2$.}
\label{FigS8}
\end{figure}
In the main text, we present the fidelity and probability of the generated mechanical states when $\omega_{M}/g_{0}=20$. In this section, as a comparison, we also study how the fidelity and probability of the mechanical states depend on the cavity field decay and the mechanical dissipation when $\omega_{M}/g_{0}=100$. These plots are shown in Fig.~\ref{FigS8}. Here the figures in the left and right boxes correspond respectively to Fig.~2 and Fig.~3 in the main text, which are plotted using $\omega_{M}/g_{0}=20$. We can see similar behaviors as those in the case of $\omega_{M}/g_{0}=20$.

\section{IV. Discussions on the experimental implementation}

\subsection{A. Measurement of the quadrature operator of the mechanical mode}

In this section, we show how to measure the quadrature operator of the mechanical mode by introducing an ancillary cavity mode in the right cavity.
The frequency of the ancillary cavity mode should be much different from that of the right cavity mode $\hat{a}_{R}$ so that the coupling between these two modes can be neglected. In this case, the measurement procedure can be described by the method proposed in~\cite{Vitali2007}.
The ancillary cavity mode couples to the mechanical oscillation via a radiation pressure mechanism and it is strongly driven by an external field so that we can linearize the dynamics of the optomechanical coupling.
When the mechanical superposition states are prepared, the photon hopping between the left and right cavities is tuned off, and then the optomechanical coupling
between the right cavity mode $\hat{a}_{R}$ and the mechanical mode $\hat{b}$ can be approximately neglected because $g_{0}$ is much smaller than other frequency scales (the free frequencies and the linearized interaction). In this case, we can only consider the linearized interaction between the ancillary cavity mode and the mechanical mode when we consider the measurement of the mechanical quadrature.

The Hamiltonian of the subsystem including the ancillary cavity mode and the mechanical mode can be written as
\begin{equation}
\hat{H}_{d}=\omega_{d}\hat{c}^{\dagger}\hat{c}+\omega_{M}\hat{b}^{\dagger}\hat{b}-g_{c}\hat{c}^{\dagger}\hat{c}(\hat{b}^{\dagger}+\hat{b})
+\Omega e^{-i\omega_{L}t}\hat{c}^{\dagger}+\Omega^{\ast}e^{i\omega_{L}t}\hat{c},
\end{equation}
where $\hat{c}$ and $\hat{b}$ are the annihilation operators of the ancillary cavity field and the
mechanical mode, respectively. The parameter $\omega_{d}$ is the frequency of the ancillary cavity mode and $g_{c}$ is the
single-photon optomechanical-coupling strength between the ancillary cavity mode and
the mechanical oscillation. In addition, the ancillary cavity mode is driven by an external field with frequency $\omega_{L}$ and driving amplitude $\Omega$.

In the rotating frame with respect to $H_{0}=\omega_{L}c^{\dagger}c$, we can obtain the Langevin equations as
\begin{subequations}
\begin{align}
\dot{\hat{c}}=&-i\Delta_{d}\hat{c}+ig_{c}\hat{c}(\hat{b}^{\dagger}+\hat{b})-i\Omega-\frac{\kappa}{2}\hat{c}+\sqrt{\kappa}\hat{c}_{\textrm{in}}, \\
\dot{\hat{b}} =&-i\omega_{M}\hat{b}+ig_{c}\hat{c}^{\dagger}\hat{c}-\frac{\gamma _{M}}{2}\hat{b}+\sqrt{\gamma_{M}}\hat{b}_{\textrm{in}},
\end{align}
\end{subequations}
where $\Delta_{d}=\omega_{d}-\omega_{L}$ is the driving detuning, $\kappa$ is the decay rate of the ancillary cavity mode. The operators $\hat{c}_{\textrm{in}}$ and $\hat{b}_{\textrm{in}}$ are the fluctuation operators of the ancillary cavity mode and the mechanical mode, respectively.

In the strong driving case, we linearize the Langevin equations by expanding the operators around their steady-state values as $\hat{c}=\varsigma_{s}+\delta \hat{c}$ and $\hat{b}=\beta_{s}+\delta \hat{b}$. Then by discarding the nonlinear terms we obtain the equations of motion for these fluctuations as
\begin{subequations}
\begin{align}
\delta \dot{\hat{c}}=&-i\Delta\delta\hat{c}+iG(\delta\hat{b}^{\dagger}+\delta\hat{b})-\frac{\kappa}{2}\delta c+\sqrt{\kappa}\hat{c}_{\textrm{in}},\\
\delta \dot{\hat{b}}=&-i\omega_{M}\delta\hat{b}+iG^{\ast}\delta\hat{c}+iG\delta\hat{c}^{\dagger}-\frac{\gamma_{M}}{2}\delta\hat{b}+\sqrt{\gamma _{M}}\hat{b}_{\text{in}},
\end{align}
\end{subequations}
where we introduce the normalized detuning $\Delta=[\Delta_{d}-g_{c}(\beta_{s}^{\ast}+\beta_{s})]$ and the linearized coupling strength $G=g_{c}\varsigma_{s}$.
In these expressions, the steady-state values $\varsigma_{s}$ and $\beta_{s}$ are determined by the equations
\begin{subequations}
\begin{align}
-i[\Delta_{d}-g_{c}(\beta^{\ast}+\beta)] \varsigma-i\Omega-\frac{\kappa}{2}\varsigma&=0,\\
-i\omega_{M}\beta+ig_{c}\varsigma^{\ast}\varsigma-\frac{\gamma_{M}}{2}\beta&=0.
\end{align}
\end{subequations}
We further assume that $\Delta=\omega_{M}\gg\vert G\vert$, then by RWA we have
\begin{subequations}
\begin{align}
\delta\dot{\hat{c}}&=-\left(i\Delta+\frac{\kappa}{2}\right)\delta
\hat{c}+iG\delta\hat{b}+\sqrt{\kappa}\hat{c}_{\text{in}},\\
\delta\dot{\hat{b}}&=-\left(i\omega_{M}+\frac{\gamma_{M}}{2}\right)\delta
\hat{b}+iG^{\ast}\delta \hat{c}+\sqrt{\gamma_{M}}\hat{b}_{\text{in}}.
\end{align}
\end{subequations}
We make a rotation
\begin{eqnarray}
\delta \hat{b} =\delta \hat{B}e^{-i\omega _{M}t},\hspace{1 cm}
\delta \hat{c} =\delta \hat{C}e^{-i\Delta t},
\end{eqnarray}
such that the high-oscillating time factors are eliminated.
Then we have
\begin{subequations}
\begin{align}
\delta \dot{\hat{C}}=&-\frac{\kappa}{2}\delta \hat{C}+iG\delta \hat{B}+\sqrt{\kappa}\hat{C}_{\textrm{in}},\\
\delta \dot{\hat{B}}=&-\frac{\gamma_{M}}{2}\delta \hat{B}+iG^{\ast}\delta \hat{C}+\sqrt{\gamma_{M}}\hat{B}_{\textrm{in}},
\end{align}
\end{subequations}
where $\hat{C}_{\textrm{in}}=\hat{c}_{\text{in}}e^{i\omega_{M}t}$ and $\hat{B}_{\textrm{in}}=\hat{b}_{\text{in}}e^{i\omega_{M}t}$.
We consider the bad-cavity case $\kappa\gg\vert G\vert$, then
the ancillary cavity mode can be eliminated adiabatically and we obtain
\begin{equation}
\delta \hat{C}=i\frac{2G}{\kappa}\delta \hat{B}+\frac{2}{\sqrt{\kappa}}\hat{C}_{\text{in}}.
\end{equation}
In terms of the input-output relation
\begin{equation}
\hat{C}_{\textrm{out}}=\sqrt{\kappa}\delta \hat{C}-\hat{C}_{\text{in}},
\end{equation}
we have
\begin{equation}
\hat{C}_{\text{out}}=i\frac{2G}{\sqrt{\kappa}}\delta \hat{B}+\hat{C}_{\text{in}},
\end{equation}
which can be further expressed as
\begin{equation}
\hat{c}_{\text{out}}=i\frac{2G}{\sqrt{\kappa}}(\hat{b}-\beta_{s})+\hat{c}_{\text{in}}.
\end{equation}
The term $\beta_{s}$ (typically with value several dozen) is a displacement in phase space, it will not change the oscillation in the probability distribution of the quadrature. By measuring the quadrature of the output field, we can observe the quantum interference and coherence effects in the superposition states. The rotation angle of the quadrature can be swept by tuning the driving amplitude $\Omega$ such that a desired phase angle $\vartheta$ is obtained by $G=|G|e^{i\vartheta}$.

\subsection{B. Discussions on experimental parameters}

\begin{table}[]
\centering
\caption{Parameters of electromechanical systems reported in references: cavity-field frequency $\omega_{c}$, cavity-field decay rate $\gamma_{c}$, cavity-field quality factor $Q_{c}=\omega_{c}/\gamma_{c}$, mechanical frequency $\omega_{M}$, mechanical decay rate $\gamma_{M}$, mechanical quality factor $Q_{M}=\omega_{M}/\gamma_{M}$, mechanical bath temperature $T_{\textrm{th}}$, thermal phonon occupation number $n_{\textrm{th}}$, single-photon optomechanical-coupling strength $g_{0}$, the ratio $\omega_{M}/\gamma_{c}$, which determines the sideband resolution, and the ratio $g_{0}/\gamma_{c}$, which determines the single-photon strong-coupling regime. Notice that if several temperatures were mentioned in a paper, here we only list the lowest one.}
\label{tablesm1}
\begin{tabular}{|c|c|c|c|c|c|c|c|c|c|c|c|}
\hline
Ref. & $\frac{\omega_{c}}{2\pi}$ (GHz) & $\frac{\gamma_{c}}{2\pi}$ (MHz) & $Q_{c}$  &  $\frac{\omega_{M}}{2\pi}$ (MHz) & $\frac{\gamma_{M}}{2\pi}$ (Hz) & $Q_{M}$  & $T_{\textrm{th}}$ (mK) &  $n_{\textrm{th}}$  &  $\frac{|g_{0}|}{2\pi}$ (Hz) & $\frac{\omega_{M}}{\gamma_{c}}$ & $\frac{g_{0}}{\gamma_{c}}$\\
\hline
\cite{Rocheleau2010} & $7.5$ & $0.6$ & $1.25\times10^{4}$ & $6.3$ & $6.3$ & $1\times10^{6}$ & $20$ & $65.6$ &   & $10.05$ &  \\
\hline
\cite{Hertzberg2010} & $5.01$ & $0.494$ & $1.01\times10^{4}$ & $5.57$ & $15$-$25$ & ($0.22$-$0.37$)$\times10^{6}$ &   &   &   &   &  \\
\hline
\cite{Teufel2011} & $7.5$ & $0.2$ & $3.75\times10^{4}$ & $10.56$ & $32$ & $3.3\times10^{5}$  & $15$ & $29.1$ & $200$ & $52.8$ & $1\times10^{-3}$\\
\hline
\cite{Teufel2011B} & $7.47$ & $0.17$ & $4.39\times10^{4}$ & $10.69$ & $30$ & $3.56\times10^{5}$ & 40 & $77.4$ & $460$ & $62.8$ & $2.7\times10^{-3}$\\
\hline
\cite{Palomaki2013A} & $7.7$ & $0.36$ & $2.14\times10^{4}$ & $10.34$ & $35$ & $2.95\times10^{5}$ & $20$ & $39.8$ & $200$ & $28.7$ & $0.56\times10^{-3}$\\
\hline
\cite{Palomaki2013B} & $7.5$ & $0.32$ & $2.34\times10^{4}$ & $10.5$ & $35$ & $3\times10^{5}$ & $15$ & $29.7$ & 200 & $32.8$ & $0.625\times10^{-3}$\\
\hline
\cite{Wollman2015} & $6.23$ & $0.45$ & $1.38\times10^{4}$ & $3.6$ & $3$ & $1.2\times10^{6}$ & $10$ & $ 57.3 $ & 36 & $ 8 $ & $0.8\times10^{-4}$\\
\hline
\cite{Pirkkalainen2015} & $6.9$ & $0.65$ & $1.06\times10^{4}$ & $13.032$ & $330$ & $3.9\times10^{4}$ & $25$ & $ 39.5 $ &   & $   $ & $ $\\
\hline
\cite{Lecocq2016} & $8.89 $ & $1.7$  & $5.23\times10^{3}$ & $14.98$ & $9.2$ & $1.63\times10^{6}$ & $30$ & $41.2$ & $145$ & $8.81$ & $0.85\times10^{-4}$\\
\hline
\cite{Lecocq2016} & $9.93$ & $2.1$ & $4.73\times10^{3}$ & $14.98$  & $9.2$ & $1.63\times10^{6}$ & $30$ & $41.2$ & $170$ & $7.13$ & $0.81\times10^{-4}$\\
\hline
\cite{Toth2016} & $4.26$ & $0.118$ & $3.64\times10^{4}$ & $5.33$  & $30$ & $1.78\times10^{5}$ & $10$  & $38.6$ & $60$ & $45.2$ & $ 5.08\times10^{-4}$\\
\hline
\end{tabular}
\end{table}
In this section, we present some detailed discussions on the experimental parameters of this scheme. For comparison purpose, in Table~\ref{tablesm1} we list some relating parameters reported in several recent experiments in electromechanics. We can see that depending on the task, the electromechanical systems can be designed to have a wide range of parameters.

(i) For superconducting resonators, the resonance frequency $\omega_{c}$ could be $2\pi\times$ ($4.26$ - $9.93$) GHz, with the decay rate $\gamma_{c}$ changing from $2\pi\times 118$ kHz to $2\pi\times 2.1$ MHz, and the quality factor $Q_{c}$ changing from $4.73\times10^{3}$ to $4.39\times10^{4}$. The decay rate $\gamma_{c}$ might be further improved because it has recently been reported in circuit-QED systems that the internal quality factors of planar superconducting resonators (with frequency at several GHz) could be above one million~\cite{Megrant2012}, then the decay rate $\gamma_{c}$ could be at the order of dozens of kilohertz.

(ii) For mechanical resonators, the resonance frequency $\omega_{M}$ could be $2\pi\times$ ($3.6$ - $14.98$) MHz, with the decay rate $\gamma_{M}$ changing between $2\pi\times$ ($3$ - $330$) Hz. The quality factor $Q_{M}$ of mechanical resonators could change from $3.9\times10^{4}$ to $1.63\times10^{6}$. In dilution refrigerators, the environmental temperature of the mechanical resonators could be at dozens of millikelvin (for example $T_{\textrm{th}}=10$ mK in Ref.~\cite{Toth2016}), and then the bath could have dozens of thermal phonons. In these references, the thermal phonon occupation number changes from $29.1$ to $77.4$.

(iii) The single-photon optomechanical-coupling strength $g_{0}$ could be $2\pi\times$ ($36$ - $460$) Hz, which needs to be enhanced for a high success probability in our scheme. Recently, people have suggested to increase the value of $g_{0}$ to around one megahertz by utilizing the nonlinearity of Josephson junction in electromechanics~\cite{Rimberg2014,Heikkila2014}.

\begin{table}[]
\centering
\caption{The parameters used in simulation of our model: cavity-field frequency $\omega_{c}$, mechanical frequency $\omega_{M}$, single-photon optomechanical-coupling strength $g_{0}$, modulation sideband parameter $n_{0}$, dimensionless photon-hopping modulation amplitude $\xi$, photon-hopping modulation frequency $\omega_{0}$, cavity-field decay rate $\gamma_{c}$, mechanical decay rate $\gamma_{M}$, thermal phonon occupation number $n_{\textrm{th}}$, state generation time $t_{d}$, maximum displacement amplitude $|\beta|_{\textrm{max}}$, and success probability $\mathcal{P}$.}
\label{table2sm}
\begin{tabular}{|c|c|c|c|}
\hline
Notation & Remarks & Scaled parameters  &Parameters  \\
\hline
$\omega_{c}$ & arbitrary & $\omega_{c}/g_{0}=$($1$ - $2$)$\times10^{4}$ & $2\pi\times$($5$ - $10$) GHz\\
\hline
$\omega_{M}$ & $\omega_{M}\gg g_{0}$ for RWA& $\omega_{M}/g_{0}=20$ & $2\pi\times 10$ MHz \\
\hline
$g_{0}$   & as the frequency scale &  $g_{0}/g_{0}=1$ & $2\pi\times 500$ kHz \\
\hline
$n_{0}$ &  1& 1 & 1\\
\hline
$\xi$ &  tunable, $g=g_{0}J_{2n_{0}}(2\xi)/2$, we choose $\xi=1.5271$&  $g/g_{0}=J_{2n_{0}}(2\xi)/2$  & $g\approx2\pi\times 121.5$ kHz  \\
\hline
$\omega_{0}$ & tunable, $\delta=\omega_{M}-2n_{0}\omega_{0}$ &    & $\delta=g$  \\
\hline
$\gamma_{c}$  &  affects probability rather than fidelity& $\gamma_{c}/g_{0}=0.05$ - $0.4$ & $2\pi\times$ ($25$ - $200$) kHz \\
\hline
$\gamma_{M}$  &  affects fidelity rather than probability& $\gamma_{M}/g_{0}=0.0001$ - $0.001$ &  $2\pi\times$ ($50$ - $500$) Hz \\
\hline
$n_{\textrm{th}}$  &  affects fidelity rather than probability&   & 30 at $15$ mK \\
\hline
$t_{d}$ &  $t_{d}\approx\pi/\delta$ & &   $4.1\mu$s \\
\hline
$|\beta|_{\textrm{max}}$ &  $|\beta|_{\textrm{max}}=2g/\delta>1$ for $|\langle\beta|-\beta\rangle|\ll1$ & &  $2$ \\
\hline
$\mathcal{P}$ & $\mathcal{P}\approx\exp(-4\pi\gamma_{c}/g_{0})$ &  &  $\mathcal{P}\approx0.0066$ - $0.53$  \\
\hline
\end{tabular}
\end{table}

Based on the above discussions, in Table~\ref{table2sm} we suggest some parameters for simulation of our state generation scheme. In our model, the involved parameters include: cavity-field frequency $\omega_{c}$, cavity-field decay rate $\gamma_{c}$, dimensionless photon-hopping modulation amplitude $\xi$, photon-hopping modulation frequency $\omega_{0}$, mechanical frequency $\omega_{M}$, mechanical decay rate $\gamma_{M}$, thermal phonon occupation number $n_{\textrm{th}}$, and single-photon optomechanical-coupling strength $g_{0}$. Below we analyze the feasibility of our scheme based on the above listed parameters.

(i) The cavity frequency could be an arbitrary number because it does not affect the quality of this approach. However, it should be taken to make sure the optomechanical Hamiltonian is justified (ensure the single-model approximation in the optomechanical model). For electromechanical systems, the cavity frequency might be $\omega_{c}=2\pi\times$ ($5$ - $10$) GHz. For high success probabilities in our state generation, the cavity-field decay rate $\gamma_{c}$ should be smaller than the single-photon optomechanical-coupling constant $g_{0}$. In our simulations, we take $\gamma_{c}=2\pi\times$ ($25$ - $200$) kHz, which is accessible with current or near-future experimental conditions. The modulation parameters $\omega_{0}$ and $\xi$ of the photon hopping are tunable. We choose $\xi=1.5271$ such that a peak value of the Bessel function $J_{2n_{0}}(2\xi)$ is obtained. We also tune the modulation frequency $\omega_{0}$ to obtain $\delta=g$ and hence a displacement amplitude $|\beta|_{\textrm{max}}=2$ is obtained.

(ii) For the mechanical resonator, we choose the resonance frequency and the decay rate as $\omega_{M}=2\pi\times 10$ MHz and $\gamma_{M}=2\pi\times$ ($50$ - $500$) Hz, which are accessible with current experimental conditions. The small $\gamma_{M}$ (under dozens of thermal phonons) makes our scheme feasible because the states are created within a finite time rather than in a long-time limit. For a bath temperature $T_{\textrm{th}}=15$ ($10$) mK, the thermal phonon occupation number is $n_{\textrm{th}}=30$ ($20$) at $\omega_{M}=2\pi\times 10$ MHz. In our simulations, due to the restriction of computational resource, we take $n_{\textrm{th}}=0$ - $10$ for a practicable dimension of truncation. Under the condition that a high quality factor can be ensured, we might choose a larger $\omega_{M}$ for a smaller $n_{\textrm{th}}$. As examples, for a bath temperature $T_{\textrm{th}}=15$ ($10$) mK, we have $n_{\textrm{th}}\approx15$ ($10$) when $\omega_{M}=2\pi\times20$ MHz.

(iii) In our method, the success probability of the state generation is approximately given by $\mathcal{P}\approx\exp(-4\pi\gamma_{c}/g_{0})$. For  an efficient state generation, the value of the ratio $g_{0}/\gamma_{c}$ should be moderately larger than $1$. For example, we have $\mathcal{P}\approx0.08$ - $0.285$ when $g_{0}/\gamma_{c}=5$ - $10$. The value of $g_{0}/\gamma_{c}$ can be increased by either increasing the coupling strength $g_{0}$ or decreasing the cavity-field decay rate $\gamma_{c}$. People have suggested to increase the value of $g_{0}$ to around one megahertz by utilizing the nonlinearity of Josephson junction in electromechanics. At the same time, as mentioned above, the cavity-field decay rate can be decreased to be dozens of kilohertz. These developments make the ratio $g_{0}/\gamma_{c}>5$ feasible. We note that a moderately large value of the ratio $g_{0}/\gamma_{c}$ (for example $g_{0}/\gamma_{c}\approx5$ - $10$) is approximately equivalent to the parameter condition for observation of clear evidences of photon blockade~\cite{Rabl2011,Liao2013} and phonon sidebands in cavity spectrum~\cite{Nunnenkamp2011,Liao2012}. In addition, it should be pointed out that though the ratio $g_{0}/\gamma_{c}$ affects the success probability, it does not affect the fidelity of the generated state. One can always generate the target states even at a low probability. In our simulations, we choose $g_{0}=2\pi\times 500$ kHz, which should be attainable with the near-future techniques.

\end{document}